# Probing condensed matter physics with magnetometry based on nitrogen-vacancy centres in diamond

Francesco Casola[1,2,*], Toeno van der Sar[1,*] and Amir Yacoby[1]

1- Department of Physics, Harvard University, 17 Oxford Street, Cambridge, MA 02138, USA

2- Harvard-Smithsonian Center for Astrophysics, 60 Garden Street, Cambridge, MA 02138, USA

* *These authors contributed equally to this work*

Correspondence to: yacoby@physics.harvard.edu

**The magnetic fields generated by spins and currents provide a unique window into the physics of correlated-electron materials and devices. Proposed only a decade ago, magnetometry based on the electron spin of nitrogen-vacancy (NV) defects in diamond is emerging as a platform that is excellently suited for probing condensed matter systems: it can be operated from cryogenic temperatures to above room temperature, has a dynamic range spanning from DC to GHz, and allows sensor–sample distances as small as a few nanometres. As such, NV magnetometry provides access to static and dynamic magnetic and electronic phenomena with nanoscale spatial resolution. Pioneering work focused on proof-of-principle demonstrations of its nanoscale imaging resolution and magnetic field sensitivity. Now, experiments are starting to probe the correlated-electron physics of magnets and superconductors and to explore the current distributions in low-dimensional materials. In this Review, we discuss the application of NV magnetometry to the exploration of condensed matter physics, focusing on its use to study static and dynamic magnetic textures, and static and dynamic current distributions.**

Introduction

Understanding the behaviour of spins and charges in materials is at the heart of condensed matter physics. In the past decades, a wide range of new materials displaying exciting physical phenomena has been discovered and explored. Examples are van der Waals materials[1], topological insulators[2,3] and complex oxide interfaces[4]. There is an intense ongoing activity focused on developing and understanding these materials and on creating new ones. The success of these efforts relies on advances in theory and material synthesis and on the development of sensitive measurement techniques. Because spins and moving charges can generate stray magnetic fields, a local and non-perturbative magnetic field sensor that can operate over broad temperature ranges is likely to be used to characterize a growing number of correlated and topological electron systems.

The spin of an elementary particle such as an electron or nucleus can be used as an atomic-scale magnetic field sensor. Thus, spin-based magnetometry techniques such as muon spectroscopy, nuclear magnetic resonance (NMR) and neutron scattering[5–7] give access to the magnetic structure of a material on the atomic scale. However, these techniques do not provide real-space imaging or sensitivity to samples with nanoscale volumes. By contrast, techniques such as magnetic force microscopy, magnetic resonance force microscopy and scanning superconducting quantum interference devices (SQUIDs)[8–12] allow real-space imaging of the magnetic fields emanating from nanoscale devices, but have a finite size and act as perturbative probes and/or in a narrow temperature range. Magnetometry based on the electron spin associated with the nitrogen-vacancy (NV) defect in diamond (FIG. 1a) combines powerful aspects of both worlds. The NV spin is an atomic-sized sensor that benefits from a large toolbox of spin manipulation techniques and can be controllably positioned within a few nanometres of the system under study. In the decade since it was first proposed[13,14] and implemented[15,16], NV magnetometry has demonstrated a combination of capabilities that sets it apart from any other magnetic-sensing technique: room-temperature single-electron[17] and nuclear[18] spin sensitivity, spatial resolution on the nanometre scale[19], operation under a broad range of temperatures (from ~1 K to above room temperature[20,21]) and magnetic fields (from zero to a few Tesla[22,23]), and non-perturbative operation. However, only in the past few years has NV magnetometry begun to explore condensed matter systems.

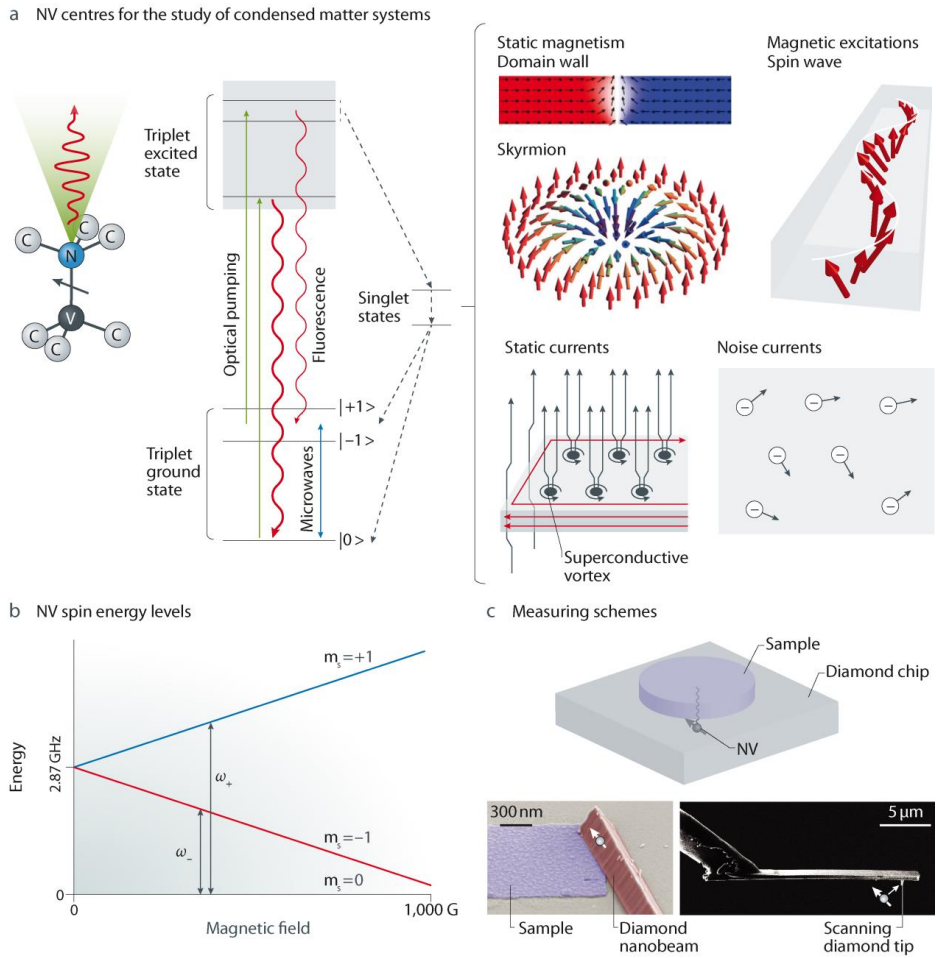

Figure 1. Probing condensed matter physics using NV magnetometry. a| The S=1 electron spin of the nitrogen-vacancy (NV) defect in diamond is a point-like magnetic field sensor that can be optically initialized and read out through its spin-dependent photoluminescence. As shown in the energy level structure, the spin is pumped into the $|0\rangle$ state by off-resonance optical excitation, the $|\pm1\rangle$ excited states can decay non-radiatively through metastable singlet states, and the ground-state spin can be manipulated by microwave excitation. The spin state can be detected through the emitted fluorescence, which is higher for the $|0\rangle$ state than for the $|\pm1\rangle$ states. In the context of probing condensed matter systems, NV magnetometry has been used to study static magnetic textures such as domain walls and skyrmions, magnetic excitations such as spin waves in ferromagnets, static current distributions such as superconducting vortices and electrical noise currents in metals. b| The energy levels of the NV spin undergo a Zeeman splitting as a function of a magnetic field applied along the NV axis. c| NV centres can be brought within a few nanometer from the sample using different approaches. Three examples are shown: the sample can be fabricated directly on diamond[36], a diamond nanostructure can be positioned on the sample[109], or an NV centre can be used in a scanning-probe configuration[21,71]. $\omega_\pm$, electron spin resonance frequencies; $m_s$, spin quantum number. Panel a is adapted from [139,140]. superconductive vortexes are courtesy of www.superconductivity.eu.

## Box 1| Measuring static fields

Here we describe elementary considerations for the use of nitrogen-vacancy (NV) centres for imaging magnetic fields generated by static magnetic textures and current distributions.

### Reconstructing a vector magnetic field by measuring a single field component

Because the NV electron spin resonance splitting is first-order sensitive to the projection of the magnetic field **B** on the NV spin quantization axis, $B_{||}$, this is the quantity typically measured in an NV magnetometry measurement[16]. It is therefore convenient to realize that the full vector field **B** can be reconstructed by measuring any of its components in a plane positioned at a distance $d$ from the sample, where $d$ is the NV–sample distance (provided this component is not parallel to the measurement plane). This results from the linear dependence of the components of $\hat{\mathbf{B}}$ in Fourier space[24,25], which follows from the fact that **B** can be expressed as the gradient of a scalar magnetostatic potential. Moreover, by measuring $B_{||}(x, y; z = d)$ we can reconstruct **B** at all distances $d + h$ through the evanescent-field analogue of Huygens' principle, a procedure known as upward propagation[24]. As an example, the out-of-plane stray field component $B_z(x, y; z = d + h)$ can be reconstructed from $B_{||}(x, y; z = d)$ using

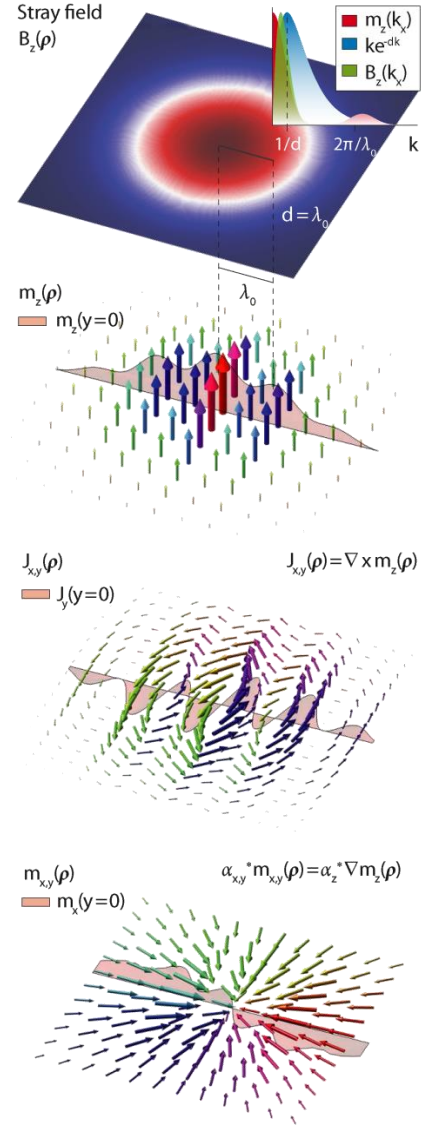

$$\hat{B}_z(\mathbf{k}; d+h) = e^{-kh} \frac{\hat{B}_{||}(\mathbf{k}; d)}{\mathbf{u}_{NV} \cdot \mathbf{u}} \qquad (1)$$

where $\mathbf{u}_{NV}$ is a unit vector in the direction of the NV quantization axis, $\mathbf{k} = (k_x, k_y)$ is the 2D wavevector and $\mathbf{u} = (-ik_x/k, -ik_y/k, 1)$. The full vector field follows from $\hat{\mathbf{B}}(\mathbf{k}) = \mathbf{u}\hat{B}_z(\mathbf{k})$. We note that reconstructing the field at distances $h < 0$ (a procedure known as downward continuation) is hampered by noise for large wavevectors[26,27].

### Magnetic field generated by planar magnetic textures and current distributions

To illustrate which sample properties may be extracted from a magnetic field measurement, we consider the field generated by a surface magnetization $\mathbf{m}(\boldsymbol{\rho}) = (\mathbf{m}_{x,y}(\boldsymbol{\rho}), m_z(\boldsymbol{\rho}))$ in the $z = 0$ plane, with $\boldsymbol{\rho} = (x, y)$. In Fourier space, the field is given by:

$$\hat{B}_i(\mathbf{k}, d) = D_{ij}^m(\mathbf{k}, d)\hat{m}_j(\mathbf{k}),$$

with $D_{ij}^m(\mathbf{k}, d) = \frac{\mu_0}{2} u_i u_j k e^{-dk}, \quad (d > 0)$ (2)

where we use the Einstein summation convention for repeated indices. The corresponding real-space expression for $B_z$ clearly illustrates its relation with the local spatial variations of the magnetization[28]:

$$B_z(\boldsymbol{\rho}, d) = -\frac{\mu_0}{2}\left[\alpha_z(\rho, d) * \nabla^2 m_z(\boldsymbol{\rho}) + \alpha_{x,y}(\rho, d) * \nabla \cdot \mathbf{m}_{x,y}(\boldsymbol{\rho})\right], \quad (d > 0). \quad (3)$$

Here, the symbol $*$ denotes a 2D convolution, and $\alpha_z(\rho, d) = \mathcal{F}_2^{-1}(e^{-kd}/k) = 1/[2\pi(\rho^2 + d^2)^{1/2}]$ and $\alpha_{x,y}(\rho, d) = \mathcal{F}_2^{-1}(e^{-kd}) = d/[2\pi(\rho^2 + d^2)^{3/2}]$ are 'resolution functions', with the NV–sample distance $d$ determining the resolving power. Using a similar formalism, we can describe the field generated by a planar current distribution. Any line current density can be expressed as the curl of an effective magnetic texture through $\mathbf{J} = \nabla \times \mathbf{m}_{\text{eff}}$. For currents confined to the $z = 0$ plane, we have $\mathbf{m}_{\text{eff}} = m_{z,\text{eff}} \hat{\mathbf{z}}$ and, consequently, $\nabla \times \mathbf{J} = -\nabla^2 m_{z,\text{eff}} \hat{\mathbf{z}}$. In Fourier space the last equation is equivalent to $\hat{m}_{z,\text{eff}} = -\varepsilon_{zlj} u_l \hat{J}_j/k$, where $\varepsilon_{zqr}$ is the Levi-Civita symbol. Using equation 2 with $m_{z,\text{eff}}$ we get

$$\hat{B}_i(\mathbf{k}, d) = D^J_{ij}(\mathbf{k}, d) \hat{J}_j(\mathbf{k}) \quad \text{with} \quad D^J_{ij}(\mathbf{k}, d) = -\frac{\mu_0}{2} u_i u_l \epsilon_{zlj} e^{-dk}, \quad (d > 0). \tag{4}$$

Similarly to equation 3, we arrive at the real-space expression:

$$B_z(\boldsymbol{\rho}, d) = \frac{\mu_0}{2} \alpha_z(\rho, d) * \big(\nabla \times \mathbf{J}(\boldsymbol{\rho})\big)_z. \tag{5}$$

**Extracting the quantities of interest**

There is a crucial difference between magnetometry of planar magnetic textures and current distributions: whereas equation 5 resembles a Poisson's equation for $m_{z,\text{eff}}$ and can be uniquely solved for $\mathbf{J}(\boldsymbol{\rho})$ from the stray field measured in a plane above the sample, this is not the case for $\mathbf{m}(\boldsymbol{\rho})$ in equation 3, because an infinite number of different magnetic textures can give rise to the same magnetic field, as illustrated in the figure. An interesting parallel can be drawn to the familiar concept of gauge freedom in electromagnetism: equation 3 resembles a Gauss' equation $B_z(\boldsymbol{\rho}, d) = -\nabla \cdot \mathbf{F}$, where $B_z$, $\mathbf{F}$, $m_z$ and $\mathbf{m}_{x,y}$ play the role of an effective charge density, electric field and scalar and vector potential, respectively[28]. Solutions can therefore be obtained via gauge-fixing, where the condition $\nabla \cdot \mathbf{m}_{x,y} = 0$ resembles the Coulomb gauge. Fixing the gauge implies fixing the helicity γ, which denotes the angle between the wavevector and the plane of rotation of the magnetic moments. For the Coulomb gauge, γ = ± π/2.

The figure shows two magnetic textures and a current distribution producing the same stray field. The out-of-plane magnetic texture $m_z(\boldsymbol{\rho})$ and the in-plane magnetic texture $\mathbf{m}_{x,y}(\boldsymbol{\rho})$ give rise to the same stray field. The two spin textures are related through a gauge transformation. The in-plane current distribution $\mathbf{J}_{x,y}(\boldsymbol{\rho})$ giving rise to the same field can be expressed as the curl of $m_z$. These plots also illustrate that fast modulations in the magnetization or current density are suppressed at distances larger than the wavenumber of the modulation; the inset in the figure shows a representation of this filtering process in Fourier space.

**Box 2 | Measuring dynamic properties**

The nitrogen-vacancy (NV) centre can act as a magnetic noise sensor and therefore allows the extraction of spectral information from a target system. Here we relate the magnetic-noise spectrum to the spin and current fluctuations in a material.

**The NV centre as a probe of a magnetic power spectral density**

For a stationary stochastic process, the power spectral density of the component *i* of the magnetic field is given by:

$$g_i(\omega) = \int_{-\infty}^{\infty} \overline{B_i(\tau) B_i(0)}\, e^{-i\omega\tau} d\tau. \tag{6}$$

where the horizontal bar denotes a thermal average[29] over the degrees of freedom of the material.

For simplicity, we use a reference frame in which the NV spin quantization axis is along *z*. Consequently, measurements of the spin coherence time of the NV are sensitive to $g_z(\omega_{dd})$, whereas relaxation measurements probe $g_x(\omega_L) + g_y(\omega_L)$, where $\omega_{dd}$ and $\omega_L$ are the dynamical-decoupling and Larmor frequencies, respectively[30,31].

**Expressing the spectral density in terms of spin–spin or current–current correlation functions**

In general, the stray fields created by fluctuating magnetic dipoles and electric currents are radiative. However, in the near-field regime defined by $d \ll c/\omega$, where *c* is the speed of light, equations 2–5 remain valid[32]. Accordingly, we can express the magnetic-field autocorrelator in equation 6 in terms of the spin–spin and current–current correlators $S_{pq}^m$ and $S_{pq}^J$, such that we arrive at:

$$g_i^{m,J}(\omega, d) = \frac{1}{4\pi^2} \int D_{ip}^{m,J}(\mathbf{k}, d) D_{iq}^{m,J}(-\mathbf{k}, d) S_{pq}^{m,J}(\mathbf{k}, \omega) d\mathbf{k}. \tag{7}$$

Note that $D_{ij}^{m,J}$ acts as a filter in Fourier space, analogous to form factors in other magnetometry techniques such as nuclear magnetic resonance[6] and neutron scattering[33]. For translationally invariant systems, the correlators are given by[29]

$$\begin{aligned} S_{pq}^m(\mathbf{k}, \omega) &= \int_{-\infty}^{\infty} \overline{\delta m_p(\mathbf{k}, \tau) \delta m_q(-\mathbf{k}, 0)}\, e^{-i\omega\tau} d\tau, \\ S_{pq}^J(\mathbf{k}, \omega) &= \int_{-\infty}^{\infty} \overline{\delta J_p(\mathbf{k}, \tau) \delta J_q(-\mathbf{k}, 0)}\, e^{-i\omega\tau} d\tau. \end{aligned} \tag{8}$$

**Expressing the correlation functions in terms of the dynamic susceptibility**

To formulate the magnetic noise emanating from a specific system such as a magnet or an electrical conductor, we use the fluctuation-dissipation theorem[29,34,35] to link the correlation functions in equation 8 to the imaginary part of the dynamical susceptibility $\chi_{pq}^{m,J\prime\prime}(\mathbf{k}, \omega)$:

$$S_{pq}^{m,J}(\mathbf{k}, \omega) = 2\hbar(n(\omega, T) + 1)\chi_{pq}^{m,J\prime\prime}(\mathbf{k}, \omega). \tag{9}$$

where $n(\omega, T) = (e^{(\hbar\omega - \mu)/kT} - 1)^{-1}$ is the Bose factor, *T* is the temperature, and $\mu$ the chemical potential. $S_{pq}$ ($\omega > 0$) describes emission processes, with the +1 term in equation 9 representing spontaneous emission into the spin or electron bath. For absorption processes (corresponding to $S_{pq}$ ($\omega < 0$)) describing energy transfers from the bath, the +1 term is absent. Examples of $\chi_{pq}\prime\prime$ for different materials are mentioned in the main text. For collinear ferromagnets, $\chi_{pq}^{m\prime\prime}$ can be described by the spin-wave dispersion[36]. For currents, $\chi_{pq}^{J\prime\prime}$ can be related to the real part of the electrical conductivity[37], as discussed in the main text. In the case of NV centres, absorption and emission processes can be associated, for example, with $0 \to \pm 1$ and $\pm 1 \to 0$ transitions, respectively. The transition rates for absorption and emission in the relaxation matrix[36] can be considered equal at room temperature, where $kT \gg \hbar\omega$.

In this Review, we describe the application of NV magnetometry to the exploration of magnetic and electron-transport phenomena in condensed-matter systems. We begin by briefly summarizing key NV properties and measurement techniques but refer the reader to existing reviews for more details[38,39,30,40–42]. This Review focuses on the material properties that can be extracted from the nanoscale magnetic fields accessible with NV magnetometry. The theoretical formalism for doing so is presented in BOX 1 (for static fields) and BOX 2 (for dynamic fields). The text is structured according to four NV magnetometry application areas (FIG. 1a). We first describe NV magnetometry studies of static magnetic textures, highlighting recent experiments that focused on determining the nature of non-collinear ferromagnetic spin textures. In the following section, we discuss the dynamic magnetic fields produced by the excitations of magnetic systems, highlighting the application of NV magnetometry to probing spin waves in ferromagnets. We then examine the fields generated by static current distributions and the first attempts to use NV defects to image current distributions in condensed matter systems. In the next section, we consider the magnetic field fluctuations created by current fluctuations in electrical conductors, which can reveal the nature of electron transport at the nanoscale. Finally, we present an outlook for future experiments.

**Magnetometry with NV centres in diamond**

The NV centre is a lattice defect in diamond (FIG. 1a) with remarkable properties. The negatively charged NV state has a $S=1$ electron spin that can be initialized through incoherent optical excitation and read out through spin-dependent photoluminescence[30,41,43]. The electron Zeeman interaction provides sensitivity to magnetic fields (FIG. 1b). Three complementary magnetic-sensing protocols[38,39,30,40–42] yield a dynamic frequency range going from DC up to ~100 GHz. First, a measurement of the NV electron spin resonance (ESR) frequencies $\omega_\pm$, typically done by sweeping the frequency of a microwave drive field and monitoring the spin-dependent photoluminuscence[43], yields the DC magnetic field $B$ through the relation $\omega_\pm = D \pm \gamma B$ (for a magnetic field oriented along the NV axis), where $D = 2.87$ GHz is the zero-field splitting and $\gamma = 2.8$ MHz G$^{-1}$ is the electron gyromagnetic ratio. Second, the transverse spin relaxation rate, which can be characterized by periodically flipping the phase of a spin superposition using microwave $\pi$-pulses (a technique known as dynamical decoupling[44–47]), is sensitive to magnetic fields at the spin-flip frequency. Third, the longitudinal spin relaxation rates, typically measured by preparing a spin eigenstate and monitoring the spin populations as a function of time (a technique known as NV relaxometry), are sensitive to the magnetic field power spectral density $g(\omega_\pm)$. This last technique makes the NV centre a field-tunable spectrometer that allows measurements of frequencies all the way up to ~100 GHz[48].

The frequency resolution of the NV centre as a magnetic field spectrometer depends on the sensing protocol. Dynamical decoupling schemes provide a frequency resolution that is typically of order KHz, limited by the decoherence time and ultimately by the lifetime of the NV spin[30,49–51]. Recently developed protocols that store information in proximal nuclear spins with longer coherence times (for example correlation spectroscopy[52]) or synchronize the repetition of a measurement sequence with a target frequency[53,54] have been able to resolve Hz-broad lines. The spectral resolution of NV relaxometry is typically limited by the NV spin dephasing rate[49], which is, roughly, of order 100 kHz.

The typical sensitivities of magnetic sensing based on single NVs range from tens of µT Hz$^{-1/2}$ for DC fields to tens of nT Hz$^{-1/2}$ for AC fields, but these numbers depend strongly on experimental parameters such as the NV centre coherence time (which can be significantly reduced by fluctuating spin baths at the diamond surface for near-surface NV centres[19,55]), the NV spin manipulation techniques, the photon collection efficiency and the use of specialized spin readout protocols[56–64]. Examples of such protocols include nuclear-spin-assisted readout [61,65–67], in which the electron spin state is mapped onto the state of a nearby nuclear spin that is subsequently readout with high fidelity, and spin-to-charge-state conversion[63], in which the NV spin state is mapped on the NV charge state, which can be readout with high fidelity. Using large ensembles[40,68,69] of NV centres can also enhance the sensitivity of NV magnetometry, bringing it to the picoTesla level.

NV centres can stably exist just a few nanometres below the diamond surface and can therefore be brought within extreme proximity to the sample. This is crucial for NV magnetometry studies of condensed matter systems, because the NV–sample distance $d$ plays a key role in determining the ability to detect and resolve magnetic properties. This role is clearly illustrated by the expression for a magnetic field generated by a planar spin texture (equation 3) or current distribution (equation 5). The ability to resolve spatial variations in the spin texture or current distribution is limited by the convolution with the so-called resolution functions $\alpha_z$ and $\alpha_{x,y}$, which are Lorentzian-like functions with a width proportional to $d$. This convolution reflects the fact that spatial variations in the magnetization or current distributions can only be resolved if $d$ is small enough. The role of the NV–sample distance can also be appreciated from the expressions for the stray field in the Fourier domain (equations 2 and 4), which show that spatial variations in magnetic textures or current distributions with wavenumbers $k \gg 1/d$ generate magnetic fields that are exponentially suppressed.

Getting NV centres close to the sample is therefore key to high-resolution spatial imaging (FIG. 1c). The most straightforward technique to achieve small NV–sample distances is to deposit the sample directly on a diamond containing shallowly implanted NV centres. The magnetic fields emanating from the sample can then be measured at the sites of these NV centres. If a material cannot be deposited directly on diamond, as is the case, for example, for materials that require epitaxial growth on a lattice-matched substrate, NV-containing diamond particles such as nanocrystals or microfabricated nanostructures[70] can be deposited on the material[13,20,59,71,28,72]. Scanning-probe magnetic imaging can be achieved in several ways: by attaching an NV-containing nanodiamond to an atomic force microscope tip[13,16,72], by microfabricating all-diamond tips hosting a shallow NV centre[20,71,73], and/or by fabricating the sample itself on a sharp tip[71,28,74,75]. Because of the importance of knowing the precise distance between the NV centre and the sample for the interpretation of the measured field data, several methods have been developed to determine $d$ with a precision of a few nanometres. One possibility is to rely on the strong dependence on $d$ of the NMR signal from protons located at the diamond surface[51] (as discussed in the section on magnetic excitations). Other methods estimate $d$ by measuring the field from a calibration sample[76,77].

**Probing static magnetic textures**

Determining the static spin configuration of a magnetic system is a central problem in condensed matter physics and is crucial for the development of magnetic devices. Powerful techniques for real-space probing of nanoscale magnetic textures are magnetic-force microscopy, X-ray magnetic circular dichroism[78] and scanning-tunnelling microscopy. NV magnetometry provides an alternative approach, which is magnetically non-perturbative and works under a wide range of magnetic fields and temperatures. However, a challenge for magnetometry is that reconstructing a magnetic texture based on stray-field measurements is an example of underconstrained inverse problem[28,24], because an infinite number of magnetic textures can give rise to the same stray field (BOX 1). Nevertheless, high-resolution, quantitative stray-field measurements can be used to reconstruct static magnetic textures under the appropriate assumptions, as discussed in this section.

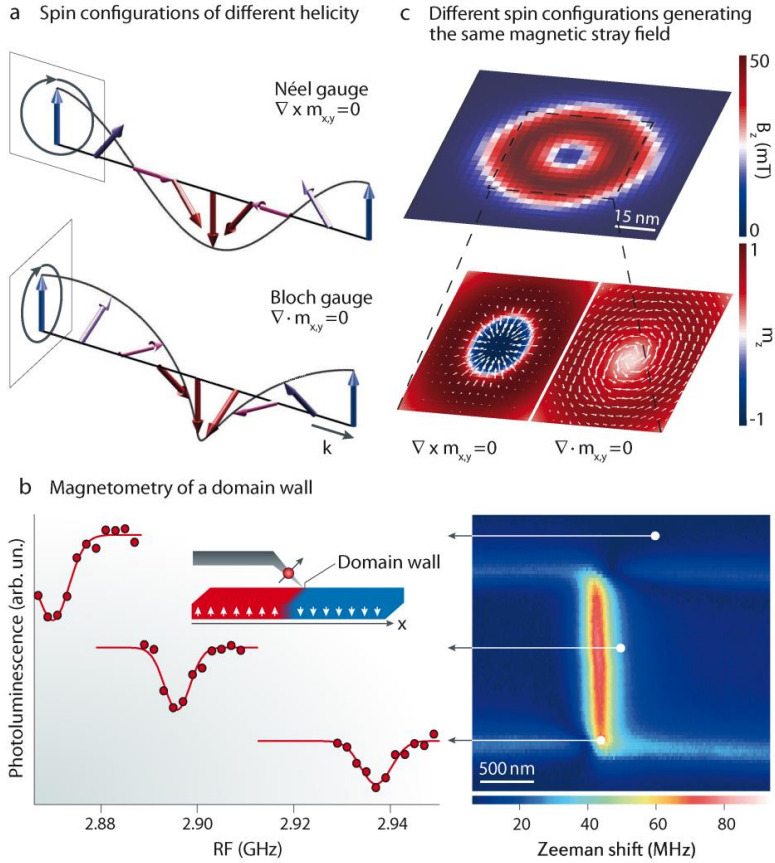

Figure 2. Imaging static magnetic textures with NV magnetometry. a| The schematic shows a Néel-like and a Bloch-like spin configuration in a 1D magnetic spiral. For $\nabla \times m_{xy} = 0$ (Néel gauge), the local moments rotate in a plane forming an angle $\gamma = 0$ or $\gamma = \pi$ with respect to the propagation vector $k$. For $\nabla \cdot m_{xy} = 0$ (Bloch gauge), the angle is $\gamma = \pm\pi/2$. b| Magnetometry of a Bloch-type domain wall in a Ta/CoFeB/MgO microbar. The right panel shows the position-dependent Zeeman shift of a single nitrogen vacancy (NV) spin in a diamond nanocrystal attached to a sharp tip that is scanned over the sample. The left panel presents individual electron spin resonance spectra for three pixels. The NV–sample distance is $d = 123$ nm.[79] c| The top image shows the simulated out-of-plane component of the magnetic stray field, $B_z$, originating from 10 layers of thin (1 nm) Co patterned in the shape of a disc with a saturation magnetization of $6 \cdot 10^5$ A m$^{-1}$. The magnetization is uniform throughout the film thickness. The distance between the NV centre and the magnetic surface is $d = 3$ nm, the scale bar is 15 nm. The bottom images display two examples out of the infinite possible spin configurations that produce the same $B_z$; the two solutions have been reconstructed from the magnetic field map imposing the Néel gauge for the image on the left, and the Bloch gauge for the image on the right, following the theory outlined in Ref. [28]. RF, radio frequency; $m_{x,y}$, in-plane magnetic moments; $m_z$, out-of-plane magnetic moment. Panel b is adapted from REF. [79]

Equation 3 explicitly shows that the stray field generated by a planar magnetic texture is determined by spatial derivatives of the local magnetization. This sensitivity to spatial variations of the magnetization motivated several NV experiments addressing the physics of domain walls in ferromagnets[28,72,79–84] [79,81–83], which connect regions with different magnetization orientations. Domain walls are well-suited for NV magnetometry because their typical widths (~10 nm)[81] match achievable NV–sample distances and domain walls thus generate easily detectable magnetic fields. A key topic of interest is to determine the nanoscale spin texture of a magnetic domain wall, which is characterized by the helicity and chirality of the domain wall[85]. The helicity describes the angle between the plane of rotation of the magnetic moments and the propagation vector[86], whereas the chirality describes the sense of rotation of the magnetic moments within this plane (FIG 2a). The determination of the spin texture of a domain wall provides fundamental insight into the magnetic anisotropies of a material and the underlying exchange energies. Furthermore, the spin texture determines the response of a domain wall to electrical currents[87,88], which is interesting for applications in race-track memories[89].

NV magnetometry studies[79,83] of thin films of X/CoFeB/MgO (where X = Ta, TaN or W ) and Pt/Co/AlO$_x$ with perpendicular magnetic anisotropy demonstrated that the nature of a domain wall can be determined under the assumption that the out-of-plane component of the magnetization profile across the domain wall is known (FIG.

2b). Following equation 3, this assumption fixes the spatial profile of the $\nabla^2 m_z$ term. Thus, different domain wall helicities, which enter through the $\nabla \cdot \boldsymbol{m}_{xy}$ term, give rise to distinct magnetic field profiles. Remarkably, even for ultra-thin magnetic films (< 1 nm), the quantitative accuracy of NV magnetometry is sufficient, under this assumption, to distinguish between domain walls of different helicity[79].

Magnetic skyrmions are nanoscale spin textures characterized by a topological number that is invariant under continuous deformations[85,90]. A skyrmion creates a region of reversed magnetization in an otherwise uniform magnet and is characterized by a helicity and a chirality, similar to domain walls[85]. Skyrmions can occur as ground states of 2D magnetic systems in the presence of chiral magnetic interactions such as the Dzyaloshinskii-Moriya interaction. They are promising candidate data bits because they can be very small (a few nanometres in diameter) and can be manipulated with relatively low currents. However, determining the spin texture of technologically interesting skyrmions in thin magnetic films is a challenging task[91] owing to the need for ~10-100 nm resolution and magnetic-field compatibility of the probing technique. Recently, this challenge was addressed using NV magnetometry[28]. Studying the stray magnetic field produced by room-temperature skyrmions in Pt/Co/Ta multilayers, it was pointed out that equation 3 resembles Gauss' equation and can therefore be solved using a procedure similar to gauge-fixing in electromagnetism (BOX 1). For instance, assuming the Bloch gauge, defined by the condition $\nabla \cdot \boldsymbol{m}_{xy} = 0$, the skyrmion helicity is fixed to $\gamma = \pm\pi/2$, because in momentum space $\boldsymbol{k} \cdot \boldsymbol{m}_{xy} = 0$ (FIG. 2a). In essence, the infinite set of spin textures compatible with a given stray field can be sorted by means of the helicity of the structure. Arguments based on the unique topology of the skyrmion were used to select the energetically stable solutions among all the possible spin textures, and the skyrmion texture could be numerically determined from stray field measurements using a steepest descent algorithm, without any additional assumptions on the texture profile[28]. An exemplary spin texture reconstruction, obtained by fixing two different gauges, is shown in FIG. 2c.

Stray-field maps can thus be used to reconstruct planar spin textures, provided that appropriate assumptions are made. The quantitative nature of the measurements and the capability of NV sensors to come extremely close to the sample are crucial assets for determining the structure of a nanoscale spin texture. The large fields generated by spin textures such as domain walls generally benefit image acquisition time, although they can also present a challenge, because a large component of the field perpendicular to the NV axis quenches the NV spin contrast (the tolerance for misalignment of the field with respect to the NV axis is less than ~10 degrees for total fields $B$ > 100 G; see Ref.[92] for details). Further applications include the imaging of magnetic hard disk write heads and hard disk drive data bits [72,93,94], the imaging and control of optically-driven 'Barkhausen' jumps of domain walls between pinning sites[82], which are crucial to understanding domain-wall dynamics, and the detection of individual magnetic nanoparticles[95]. Finally, it is worth noting that long-wavelength periodic modulations can be spotted even in materials featuring spin textures that would otherwise produce no stray field. An example is the weak ferromagnetism[96] in classical antiferromagnets with chiral anisotropies. In these systems, a small canting angle between otherwise fully antiferromagnetically coupled moments can produce a stray field that can be detected with an NV magnetometer[97].

**Probing magnetic excitations**

Correlated electron systems support a wealth of magnetic excitations, ranging from spin waves to exotic fractional excitations in low-dimensional or geometrically frustrated spin systems[98]. The spectral function $S^m(\boldsymbol{k},\omega)$ of a spin system can be probed by measuring the magnetic field noise in the vicinity of the surface of the sample (equation 7). Quantities such as the lifetime or coherence time of the NV centre are related to transition rates between opposite states $|i\rangle$ and $|f\rangle$ on the Bloch sphere. Such transitions are faster the larger the noise spectral density $g(\omega)$ at the probed frequency $\omega$ (BOX 2). The coupling between the spectral function of the material and the noise spectral density is described in momentum space by a characteristic form factor, $D(k)$. Similar form factors play a major role in other kinds of spin-based magnetometers such as neutron scattering[99], nuclear magnetic resonance[6] and muon spectroscopy[5]. Form factors are crucial for understanding the sensitivity of a technique to spatial spin–spin fluctuations. For NV magnetometry, $D(k)$ shows a peak[36] at $k = 1/d$.

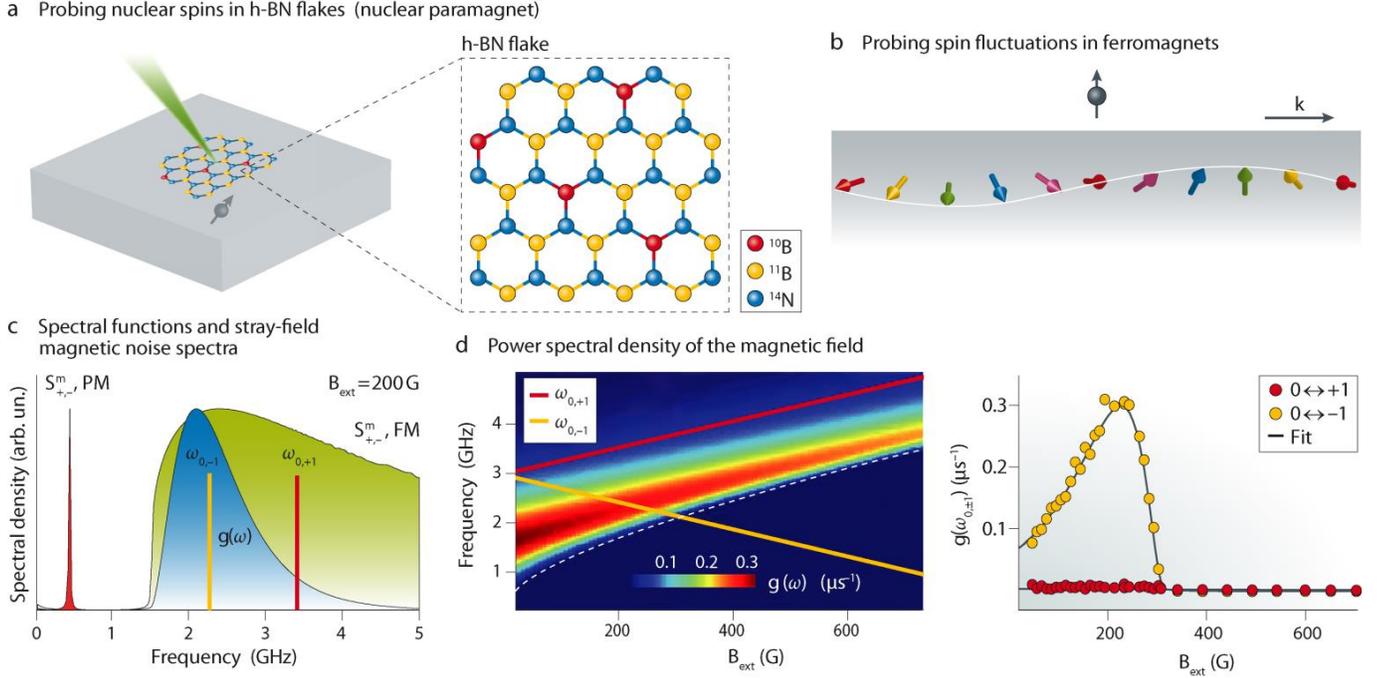

Figure 3. Probing thermally excited spin systems. a| The schematic shows a flake of hexagonal boron nitride (h-BN) on top of a diamond containing shallow nitrogen-vacancy (NV) centres[106]. The NV centres were used to probe the nuclear spins in h-BN flakes down to a thickness of one monolayer.[106] b| Schematic image showing an NV centre probing a long-wavelength spin fluctuation with momentum $k$ in a collinear ferromagnet. c| Schematic illustration of the spectral functions and associated stray-field magnetic noise spectra for a paramagnetic spin bath (PM) and a correlated-electron spin system (FM). The red peak is the transverse spectral function $S_{+,-}^m(\omega) = S_{x,x}^m(\omega) = S_{y,y}^m(\omega)$ for uncorrelated spins, showing a simple peak at the Larmor frequency for a bias field of 200 G. The green-shaded curve depicts the spectral function $S_{y,y}^m(\omega)$ for spins in a 20 nm magnetic yttrium iron garnet (YIG) film[109] ($y$ is the in-plane direction transverse to the magnetization). The energy minimum of the spectral function in this case does not coincide with that for paramagnetic spins owing to dipolar energies. The blue-shaded curve shows the stray field noise resulting from $S_{y,y}^m(k, \omega)$ after the filter functions in equation 7 have been used. The red and yellow lines represent the transition frequencies of the NV centre. d| The colour map shows the calculated power spectral density $g(\omega)$ of the magnetic field created by thermal spin waves in the YIG film at a distance $d$ = 110 nm (the thickness of the film is 20 nm). The white dashed line indicates the bottom of the spin-wave band, which coincides with the ferromagnetic resonance. The measurements of the NV spin relaxation rates shown in the right panel quantify the power spectral density along the NV electron spin resonance frequencies[109]. Panel a is adapted from REF. [106]. The right part of panel d is adapted from REF. [109].

Insight into the spectral properties of a spin system can be gained from the fluctuation-dissipation theorem, which in linear response theory relates the statistical thermal fluctuations to the imaginary part of the dynamical magnetic susceptibility $\chi^{m,\prime\prime}(k, \omega)$[34,29]. In this section, we describe experiments that have probed thermally fluctuating as well as driven spin systems. We start by discussing a thermally fluctuating paramagnetic spin bath before moving on to ferromagnets and then to the magnetic fields that are generated by ferromagnetic systems subjected to microwave drive fields.

As a simple example of the relation between thermal spin fluctuations and the associated magnetic noise spectrum at the NV site, we consider an isotropic 2D paramagnetic spin system (for example, a nuclear spin bath). The noise spectrum is given by equation 7. In this case, $S^m$ contains only one distinct transverse matrix element. The absence of spatial spin–spin correlations implies that $S^m$ is momentum-independent and can therefore be taken out of the integral in equation 7. Consequently, the perpendicular-to-plane magnetic noise spectrum from a 2D paramagnetic layer is simply given[51] by $g_z(\omega, d) = 3\mu_0^2/(64\pi d^4) \cdot (2k_B T)/\omega \cdot \chi^{m,\prime\prime}(\omega)$ for an external magnetic field oriented along $z$ and for $kT \gg \hbar\omega$. Here, $\chi^{m,\prime\prime}$ is the dissipative part of the transverse dynamical susceptibility (equation 9), which has a peak at the nuclear Larmor frequency. The power contained in this peak can be obtained from the longitudinal (Curie–Weiss) susceptibility $\chi'(\omega = 0)$ through the Kramers-Kronig[29] relation $\chi'(\omega = 0) = \int 2\,\chi^{m,\prime\prime}(\omega)/\omega\, d\omega = g^2\mu_N^2/4k_B T \cdot \rho$ (for a spin ½ system), where $g$ is the nuclear g-factor and $\mu_N$ the nuclear magnetic moment. This way it is possible to compute the root mean square of the magnetic field noise for a semi-infinite layer of nuclear spins $\overline{B_{z,\text{rms}}^2} = \int \int_d^\infty g_z(\omega, z)\, dz\, d\omega$. Such noise scales as $\sim 1/d^3$ and has been measured[51] to be $\sim$(300 nT)$^2$ for a proton layer positioned 10 nm from the NV centre.

The thermal fluctuations of several nuclear spin baths have been probed using NV magnetometry, typically by measuring the NV spin decoherence rate while flipping the NV spin in sync with the nuclear spin dynamics[51,62,100–105]. It is interesting to note that the signal is dominated by nuclear spins within a volume set by the distance from the NV centre, which can be as small as a few nanometres. A recent example in condensed matter is the NV-based detection of the quadrupolar resonance of boron spins in a monolayer of boron nitride[106] (FIG. 3a).

Moving to correlated-electron spin systems, ferromagnets constitute an excellent testbed for NV magnetometry, because their correlated nature leads to spin–spin correlations on length scales that are readily accessible with an NV sensor (FIG. 3b). In particular, the long-wavelength fluctuations perpendicular to the quantization axis of the ferromagnet can be tuned into resonance with the NV centre transition frequencies by means of an applied magnetic field[36]. We can therefore express the relevant transverse fluctuations $S_\perp^m(\mathbf{k}, \omega)$ in terms of the dissipative part of the dynamic susceptibility of the ferromagnet $\chi_\perp''(\omega, \mathbf{k})$ (BOX 2). As an example, we consider an isotropic 2D ferromagnetic film that is homogeneously magnetized by an in-plane external field. If we assume that out-of-plane fluctuations are suppressed by the demagnetizing fields they create, $\chi_\perp''(\omega, \mathbf{k})$ has only one distinct element. This element describes the in-plane fluctuations generated by spin-wave excitations and can be approximated by a peak centred at the spin-wave dispersion $\omega_k = D_s k^2$, where $D_s$ is the spin stiffness; thus, $\chi_\perp''(\omega, \mathbf{k}) \sim \delta(\omega - \omega_k)$ [107,108]. The magnetic noise spectrum at the NV site is obtained by substituting $\chi_\perp''(\omega, \mathbf{k})$ into equations 9 and 7. Unlike in the paramagnetic spin system, the magnetic noise spectrum of the ferromagnet is broad in frequency owing to the continuous nature of the spin-wave band, and its power spectral density decreases above frequencies corresponding to spin-wave excitations with wavenumbers larger than the NV–film distance (FIG. 3c)[36]. Measurements of relaxation rates of NV spins above permalloy[36] and yttrium iron garnet (YIG) films[109] (FIG. 3d) are excellently reproduced by models based on thermally excited spin waves in these films, demonstrating that the GHz magnetic fields emanating from magnetic samples provide a unique window into their magnetic excitation spectra.

Several NV magnetometry experiments have started to explore spin-wave physics in ferromagnets excited with microwaves[36,109–114]. In REF.[110] it was shown that driving the ferromagnetic resonance of a YIG film reduced the photoluminescence of non-resonant NVs located ~100 nm from the film. This phenomenon provides a convenient new technique for broadband detection of spin-wave resonances, which does not rely on matched ESR and ferromagnetic resonance (FMR) frequencies. This technique was used to study the rich ferromagnetic resonance modes of µm-thick YIG films[111]. Further experiments[109,112] unravelled the mechanism underlying this FMR-drive-induced noise, showing that FMR driving generates high-energy spin waves that can be resonant with the NV ESR frequency, thereby inducing NV spin relaxation and suppressing the NV photoluminescence. These measurements constitute the first steps towards characterizing spin-wave spectra with NV magnetometry.

Measurements of the magnetic fluctuations generated by a spin system can be used to extract the chemical potential of a spin-wave bath coupled to the NV centre, as was demonstrated in REF.[109], in which magnetic noise measurements were linked for the first time to a key spin transport quantity. Studying spin waves in a 20-nm-thick YIG film, it was also found that driving the FMR provides an efficient method for increasing the spin chemical potential. A comparison of the drive-power dependence of this process to a two-fluid theory of the coupling between the FMR and the thermal spin-wave bath yielded an experimental estimate of the 'thermomagnonic torque' between the FMR and the thermal spin-wave bath. This quantity is interesting for the growing field of spin caloritronics[115], which focuses on the interaction between heat and spin waves.

The magnetic fields generated by spin waves or by other forms of collective spin dynamics, such as domain-wall motion, could be interesting for technological applications. In particular, such collective spin dynamics can locally amplify externally applied magnetic drive fields and may mediate coupling between spin qubits[36,113,114]. NV magnetometry measurements showed that the fields generated by such spin dynamics can easily exceed the external drive field[36,113,114], leading to large locally generated microwave magnetic fields. Furthermore, if a propagating spin-wave mode is excited by an external drive field, the mode may serve as a bus to deliver the field to remote locations[114], providing, for example, enhanced remote control of spin qubits such as the NV spin.

**Static current distributions**

We now turn to the application of NV magnetometry to the characterization of the magnetic fields generated by static electrical current distributions. In mesoscopic condensed matter systems, the spatial distribution of electrical currents plays a prominent role in some of the most intriguing known physics phenomena. Examples are edge currents in quantum Hall systems and vortices in superconductors; van der Waals materials such as graphene also host a range of phenomena associated with interesting current distributions, such as the Snell's law for electrons[116], viscous electron flow[117], electron focusing[118] and Klein tunneling[119]. In this section, we describe how current distributions can be reconstructed from stray field measurements, providing a way to realize spatially resolved transport experiments, and we summarize the first steps of NV magnetometry in this field.

An advantage of studying currents is that 2D current distributions can be uniquely reconstructed from the measurement of any component of the stray field in a plane above the sample (BOX 1). Inverting the convolution of the current distribution with the resolution function $\alpha_z$ in equation 5 (BOX 1) is a simple operation in Fourier space. However, it should be noted that this procedure is an example of a downward continuation calculation, which is hampered by noise at high spatial frequencies[24,26,27]. To favour smooth solutions in the inversion, classical regularization methods such as the Tikhonov method[120] can be used. One possibility is to start from the real-space version of equations 4 and 5 and minimize a cost function[120] that includes regularization terms proportional to $J_j^2$. Inversion procedures have been used to reconstruct the current flow in integrated circuits[27], nanowires[26] and graphene[121]. Measurements of the field generated by the flow of current around defects in a graphene flake (FIG. 4a) nicely illustrate the sensitivity of the out-of-plane stray field component to the local vorticity $\nabla \times \mathbf{J}$, as described by equation 5. These measurements represent the first steps towards probing current distributions in condensed matter systems with NV magnetometry.

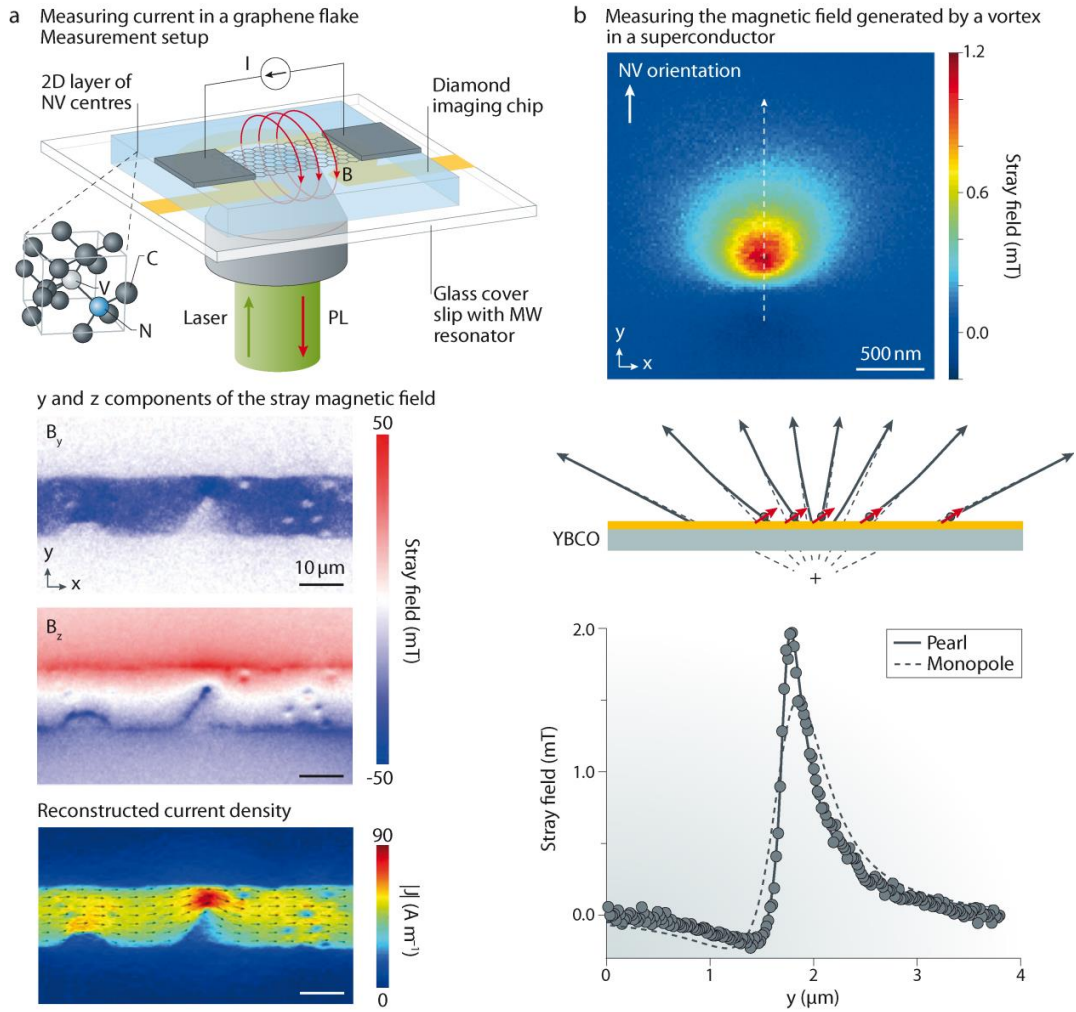

**Figure 4. NV magnetometry of static current patterns. a|** Measurements of the current flowing in a graphene flake. A schematic depiction of the measurement setup is shown in the top panel: a graphene flake is deposited on a diamond chip containing a near-surface layer of nitrogen vacancy (NV) centres[121]. The NV–graphene distance is ~20 nm and a gold stripline is used for microwave (MW) delivery.[121] The middle panel shows the measured in-plane (*y*) and out-of-plane components (*z*) of the stray magnetic field generated by a 0.8 mA current flowing in the graphene flake[121]. The bottom panel displays the magnitude of the current density in the graphene flake, reconstructed from the measured components of the magnetic field. **b|** Cryogenic scanning-NV magnetometry measurement of the stray field generated by a vortex in a 100-nm-thick superconducting YBa$_2$Cu$_3$O$_{7-\delta}$ (YBCO) film. The projection of the field on the NV quantization axis is shown in the top panel. The NV–superconductor distance is ~70 nm. The middle panel shows a sketch of the magnetic field generated by a vortex within the monopole (dashed lines) and Pearl (solid lines) approximation. The orange line shows the location of the line trace indicated by the white dashed line in the top panel; the NV orientation is also shown. In the bottom panel, the stray field measured along the white dashed line in the top panel (dots) is compared with the stray field calculated within the monopole (dashed lines) and Pearl (solid lines) approximations[21]. *I*, current; PL, photoluminescence. Panel a is adapted from REF. [121]. Panel b is adapted from REF. [21].

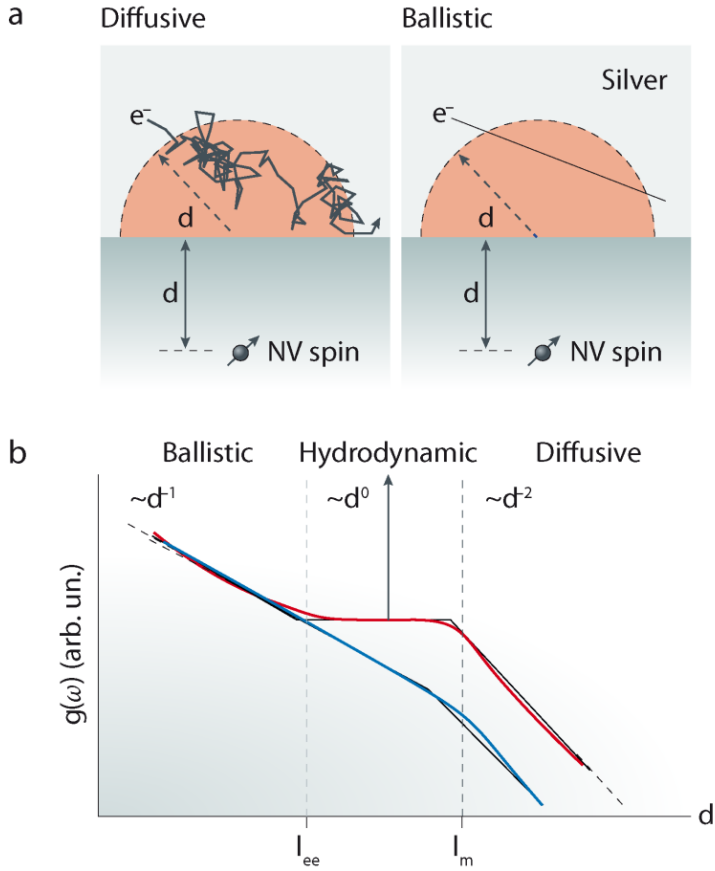

**Figure 5. Magnetic noise generated by current fluctuations in an electron liquid. a|** Schematic illustration showing the diffusive and ballistic electron motion in a metal[126]. These regimes are characterized by $d \gg l_m$ and $d \ll l_m$, respectively, where $l_m$ is the electron mean free path and $d$ the distance between the nitrogen-vacancy (NV) centre and the sample. **b|** The plot shows the theoretically expected dependence on distance of the magnetic noise generated by a metallic film in the 2D limit, after REF.[37]. The curves show, schematically, the magnetic noise as a function of the sensor–sample distance, $d$, in various transport regimes; $l_m$ and $l_{ee}$ are the mean free paths due to, respectively, extrinsic electron scattering (caused by phonons and/or impurities) and intrinsic (interparticle) electron scattering. The blue curve describes the situation when $l_{ee} > l_m$ and the hydrodynamic regime is absent. $g(\omega)$, power spectral density. Panel a is adapted from REF. [126]. Panel b is adapted from REF. [37].

In recent experiments, NV-based magnetic imaging was extended to cryogenic temperatures and used to investigate vortices in type II superconductors[20,21,122]. Superconducting vortices are small regions of size $\xi$ (the coherence length) in which the superconducting order parameter is suppressed, plus a larger region of size $\lambda$ (the penetration depth) in which a persistent current circulates [123]. When determining the stray field **B**, for distances $d \gg \lambda$ the microscopic details of the vortex are irrelevant and the vortex stray field resembles that of a magnetic monopole[124], $\mathbf{B} = \Phi_0/2\pi r^2\, \hat{\mathbf{r}}$, where $\Phi_0$ is the flux quantum and $r$ is the distance to the monopole. The radial nature of the stray field is a consequence of the Meissner effect[124], which prevents the field lines from closing on themselves. High-resolution magnetic field maps taken ~70 nm above a film of $YBa_2Cu_3O_{7-\delta}$ (FIG. 4b) showed[21] that the description in terms of monopoles breaks down for films with a thickness $t \ll \lambda$, and that for these films a correction, called the Pearl correction,[125] has to be included to correctly describe the vortex stray field (FIG. 4b). The resulting quantitative stray-field measurements allowed the extraction of the London penetration depth. These experiments are the first examples of high-resolution, quantitative NV-based magnetic imaging at low temperature, and open the way for the measurement of a wealth of interesting low-temperature physics in quantum systems and devices.

**Dynamics of the electron liquid**

Similar to static current patterns producing static magnetic fields, thermal currents produce magnetic noise fields that can be characterized using NV magnetometry. In this section we discuss how the noise generated by electronic currents can provide information about the energy-momentum dependence of the electrical conductivity of a material.

The magnetic fields generated by thermal noise currents can be expressed in terms of the dissipative part of the dynamical current–current susceptibility, $\chi''^{,J}$, using the fluctuation-dissipation theorem[37] (equation 9). This susceptibility describes[29] the linear change in the expectation value of the current density $\delta J(r, t)$ due to a term that couples to it in the Hamiltonian. Because the magnetic vector potential $A$ couples to the current density, it is possible to define $\delta J(k, \omega) = \chi^J(k,\omega)\delta A(k,\omega)$. Because the electrical conductivity, σ, is defined by $\delta J(k, \omega) = \sigma(k,\omega)\delta E(k,\omega)$, where $E$ is the electric field, and because we can use a gauge[37] in which $E = i\omega A$, we can conclude that $\chi''^{,J}(k, \omega) = \omega\sigma'(k,\omega)$. Therefore, measurements of the stray magnetic noise from a conducting material can provide insight into the energy-momentum dependence of the real part of the electrical conductivity. For isotropic media, $\chi^J$ has only two distinct matrix elements, which correspond to longitudinal and transverse current fluctuations (that is, fluctuations with $k \times J = 0$ and $k \cdot J = 0$, respectively). Furthermore, longitudinal current fluctuations are suppressed because of corresponding screening charge fluctuations[37]. The magnetic noise is thus expressible in terms of the real part of the transverse conductivity, $\sigma'^T$.

As an example, by combining equations 7, 8 and 9, the power spectral density of the stray-field magnetic noise perpendicular to a 2D conductor is given by $g_{2D}{}^d(\omega,d) = 2k_B T \Sigma_{i=x,y} \int \sigma'^T(k,\omega)\ D_{zi}^J(k,d)\ D_{zi}^J(-k,d)\ d^2k$ for $k_B T \gg \hbar\omega$. The $k^2 e^{-2kd}$ term in the form factor $D^J(k,d)\ D^J(-k,d)$ shows that the noise is once again dominated by correlations on the scale of the NV–sample distance. If $d \gg l_m$, where $l_m$ is the electron mean free path, the noise is mainly associated with diffusive electron transport[37,126] (FIG. 5a). In this case, $\sigma'^T(k,\omega)$ can be replaced by the Drude conductivity $\sigma(k\to 0, \omega) = ne^2\tau/m(1 - i\omega\tau)$, where $n, m, e$ and $\tau$ are the electron density, mass, charge and scattering time, respectively; this leads to a characteristic scaling of the power spectral density $g_{2D}{}^d \propto 1/d^2$.

In the opposite limit of ballistic transport, for which $d \ll l_m$ (FIG. 5a), the correlation time of the current–current fluctuations probed by the NV sensor is limited[126] by the time it takes an electron to ballistically traverse a distance $d$, thus it is possible to use the replacement $\tau \approx d/v_F$, with $v_F$ the Fermi velocity, into the expression for the Drude conductivity to get $g_{2D}{}^d \propto 1/d$. For the hydrodynamic limit[117], in which electron–electron interactions dominate electron scattering, the scaling of $\sigma'^T(k)$ was derived[37] by finding solutions to the Navier–Stokes equations for incompressible fluids. The scaling of the noise with $d$ thus provides insight into the nature of electron transport on the scale of $d$ (FIG. 5b). The possibility of selecting different $k$-sectors of $\sigma'^T(k)$ simply by tuning the distance between sample and sensor is ultimately related with the fact that the momentum filter $D^J(k,d)$ contains the Laplace transform kernel $e^{-kd}$. A set of measurements of $g(\omega)$ at different $d$ can therefore be used to get information about the conductivity $\sigma'^T(k)$ over a range of wave vectors[37].

The first NV experiments probing thermal currents[126] measured the magnetic noise from silver films as a function of $d$. The noise from polycrystalline silver films was found to scale with $d$ in the characteristic way of diffusive electron transport, whereas single-crystal films showed a deviation from this scaling behaviour at small $d$ that was attributed to ballistic electron transport. This method of locally characterizing the conductivity opens the way to the study of a wide range of systems with interesting and potentially gate-tunable electron transport behaviours, such as quantum Hall systems and/or van der Waals materials.

**Summary and outlook**

In this Review, we described how NV magnetometry has started to be used to explore the rich world of condensed matter physics. A powerful aspect of NV sensors is that they are point-like, so that the material properties accessible to NV magnetometry can be identified simply by deriving expressions for stray magnetic fields. We expressed these fields in terms of the properties of magnetic and electronic systems by formulating the form factors of the technique, thereby following an approach similar to that used in other spin-based magnetometry methods[5,6,33]. This approach illustrates that NV magnetometry is most effective in probing static or dynamic magnetic and electronic phenomena that spatially vary on the scale of the NV–sample distance. Furthermore, we discussed how magnetometry is, in general, well suited for unravelling static mesoscopic planar current distributions, as these can be uniquely reconstructed from stray field measurements. By contrast, uniquely reconstructing a spin texture from field measurements requires additional assumptions or knowledge of some system properties.

**Outlook**

Many exciting opportunities for NV magnetometry applied to condensed matter systems lie beyond the applications described in this Review. Imaging the nanoscale spin textures and excitations of more exotic correlated-electron systems, such as complex-oxide interfaces, multiferroics and recently discovered monolayer van-der-Waals magnets[127,128] is likely to yield many interesting results. We anticipate a growing range of applications for NV centres for probing spin-wave physics, including nanoscale imaging of spin-wave transport and local, quantitative characterization of spin transport parameters. The ability to locally probe the susceptibility of a material to microwave magnetic fields with amplitude- and phase-sensitivity[76,36,26] may provide the opportunity to perform measurements similar to electrical microwave impedance microscopy[129]. The broad temperature operability of NV sensors combined with its ability to measure temperature *in-situ*[130] makes them well suited for probing magnetic phase transitions[131], spin caloritronic phenomena[115] and, in general, the many magnetic systems with cryogenic Curie temperatures and the low-temperature physics of quantum materials and devices[20,21,132]. Recent work has also presented solutions to extend NV magnetometry to applied fields of a few Teslas, a range in which the NV transition frequencies begin to approach the ~100 GHz limit[22,23].

Looking further ahead, there are theoretically proposed protocols that would allow measurements of real-space two-point spin–spin correlation functions,[133] and the use of spin-wave excitations to mediate interactions between distant NV centres was discussed[134]. In addition, the spatial resolution of paramagnetic electron and nuclear spins can be considerably increased through, for example, the use of magnetic field gradients similar to those used for MRI[19,60]. This may enable the extraction of lattice-scale information about the electronic structure of a material. Finally, beyond magnetic field sensing, the NV centre can be used as a local probe of thermal gradients [59] and electric fields[135,136], and its optical lifetime can be used to quantify the local photonic densities of states [137,138].

**References**

1. Geim, A. K. & Grigorieva, I. V. Van der Waals heterostructures. *Nature* **499,** 419–425 (2013).

2. Hasan, M. Z. & Kane, C. L. *Colloquium* : Topological insulators. *Rev. Mod. Phys.* **82,** 3045–3067 (2010).

3. Soumyanarayanan, A., Reyren, N., Fert, A. & Panagopoulos, C. Emergent phenomena induced by spin-orbit coupling at surfaces and interfaces. *Nature* **539,** 509–517 (2016).

4. Hwang, H. Y., Iwasa, Y., Kawasaki, M., Keimer, B., Nagaosa, N. & Tokura, Y. Emergent phenomena at oxide interfaces. *Nat. Mater.* **11,** 103–113 (2012).

5. Blundell, S. J. Spin-polarized muons in condensed matter physics. *Contemp. Phys.* **40,** 175–192 (1999).

6. Walstedt, R. E. *The NMR Probe of High-Tc Materials*. *Springer Tracts Mod. Phys.* **231,** (2008).

7. Bramwell, S. T. & Keimer, B. Neutron scattering from quantum condensed matter. *Nat. Mater.* **13,** 763–767 (2014).

8. Embon, L., Anahory, Y., Suhov, A., Halbertal, D., Cuppens, J., Yakovenko, A., Uri, A., Myasoedov, Y., Rappaport, M. L., Huber, M. E., Gurevich, A. & Zeldov, E. Probing dynamics and pinning of single vortices in superconductors at nanometer scales. *Sci. Rep.* **5,** 7598 (2015).

9. Lee, I., Obukhov, Y., Xiang, G., Hauser, A., Yang, F., Banerjee, P., Pelekhov, D. V. & Hammel, P. C. Nanoscale scanning probe ferromagnetic resonance imaging using localized modes. *Nature* **466,** 845–848 (2010).

10. Vasyukov, D., Anahory, Y., Embon, L., Halbertal, D., Cuppens, J., Neeman, L., Finkler, A., Segev, Y., Myasoedov, Y., Rappaport, M. L., Huber, M. E. & Zeldov, E. A scanning superconducting quantum interference device with single electron spin sensitivity. *Nat. Nanotechnol.* **8,** 639–644 (2013).

11. Nowack, K. C., Spanton, E. M., Baenninger, M., König, M., Kirtley, J. R., Kalisky, B., Ames, C., Leubner, P., Brüne, C., Buhmann, H., Molenkamp, L. W., Goldhaber-Gordon, D. & Moler, K. A. Imaging currents in HgTe quantum wells in the quantum spin Hall regime. *Nat. Mater.* **12,** 787–791 (2013).

12. Spinelli, A., Bryant, B., Delgado, F., Fernández-Rossier, J. & Otte, A. F. Imaging of spin waves in atomically designed nanomagnets. *Nat. Mater.* **13,** 782–785 (2014).

13. Degen, C. L. Scanning magnetic field microscope with a diamond single-spin sensor. *Appl. Phys. Lett.* **92,** 243111 (2008).

14. Taylor, J. M., Cappellaro, P., Childress, L., Jiang, L., Budker, D., Hemmer, P. R., Yacoby, A., Walsworth, R. & Lukin, M. D. High-sensitivity diamond magnetometer with nanoscale resolution. *Nat. Phys.* **4,** 810–816 (2008).

15. Maze, J. R., Stanwix, P. L., Hodges, J. S., Hong, S., Taylor, J. M., Cappellaro, P., Jiang, L., Dutt, M. V. G., Togan, E., Zibrov, A. S., Yacoby, A., Walsworth, R. L. & Lukin, M. D. Nanoscale magnetic sensing with an individual electronic spin in diamond. *Nature* **455,** 644–647 (2008).

16. Balasubramanian, G., Chan, I. Y., Kolesov, R., Al-Hmoud, M., Tisler, J., Shin, C., Kim, C., Wojcik, A., Hemmer, P. R., Krueger, A., Hanke, T., Leitenstorfer, A., Bratschitsch, R., Jelezko, F. & Wrachtrup, J. Nanoscale imaging magnetometry with diamond spins under ambient conditions. *Nature* **455,** 648–651 (2008).

17. Grinolds, M. S., Hong, S., Maletinsky, P., Luan, L., Lukin, M. D., Walsworth, R. L. & Yacoby, A. Nanoscale magnetic imaging of a single electron spin under ambient conditions. *Nat. Phys.* **9,** 215–219 (2013).

18. Sushkov, a. O., Lovchinsky, I., Chisholm, N., Walsworth, R. L., Park, H. & Lukin, M. D. Magnetic Resonance Detection of Individual Proton Spins Using Quantum Reporters. *Phys. Rev. Lett.* **113,** 197601 (2014).

19. Grinolds, M. S., Warner, M., De Greve, K., Dovzhenko, Y., Thiel, L., Walsworth, R. L., Hong, S., Maletinsky, P. & Yacoby, A. Subnanometre resolution in three-dimensional magnetic resonance imaging of individual dark spins. *Nat. Nanotechnol.* **9,** 279–284 (2014).

20. Pelliccione, M., Jenkins, A., Ovartchaiyapong, P., Reetz, C., Emmanouilidou, E., Ni, N. & Bleszynski Jayich, A. C. Scanned probe imaging of nanoscale magnetism at cryogenic temperatures with a single-spin quantum sensor. *Nat. Nanotechnol.* **11,** 700–705 (2016).

21. Thiel, L., Rohner, D., Ganzhorn, M., Appel, P., Neu, E., Müller, B., Kleiner, R., Koelle, D. & Maletinsky, P. Quantitative nanoscale vortex imaging using a cryogenic quantum magnetometer. *Nat. Nanotechnol.* **11,** 677–681 (2016).

22. Stepanov, V., Cho, F. H., Abeywardana, C. & Takahashi, S. High-frequency and high-field optically detected magnetic resonance of nitrogen-vacancy centers in diamond. *Appl. Phys. Lett.* **106,** 063111 (2015).

23. Aslam, N., Pfender, M., Stöhr, R., Neumann, P., Scheffler, M., Sumiya, H., Abe, H., Onoda, S., Ohshima, T., Isoya, J. & Wrachtrup, J. Single spin optically detected magnetic resonance with 60–90 GHz (E-band) microwave resonators. *Rev. Sci. Instrum.* **86,** 064704 (2015).

24. Blakely, R. J. *Potential Theory in Gravity and Magnetic Applications*. *Program* (1995). doi:10.1017/CBO9780511549816

25. Lima, E. A. & Weiss, B. P. Obtaining vector magnetic field maps from single-component measurements of geological samples. *J. Geophys. Res.* **114,** B06102 (2009).

26. Chang, K., Eichler, A. & Degen, C. L. Nanoscale imaging of current density with a single-spin magnetometer. (2016). at <http://arxiv.org/abs/1609.09644>

27. Nowodzinski, A., Chipaux, M., Toraille, L., Jacques, V., Roch, J.-F. & Debuisschert, T. Nitrogen-Vacancy centers in diamond for current imaging at the redistributive layer level of Integrated Circuits. *Microelectron. Reliab.* **55,** 1549–1553 (2015).

28. Dovzhenko, Y., Casola, F., Schlotter, S., Zhou, T., Büttner, F., Walsworth, R. L., Beach, G. S. D. & Yacoby, A. Imaging the Spin Texture of a Skyrmion Under Ambient Conditions Using an Atomic-Sized Sensor. *https://arxiv.org/abs/1611.00673* 1–15 (2016).

29. Schwabl, F. *Advanced Quantum Mechanics*. (Springer Berlin Heidelberg, 2008). doi:10.1007/978-3-540-85062-5

30. Rondin, L., Tetienne, J.-P., Hingant, T., Roch, J.-F., Maletinsky, P. & Jacques, V. Magnetometry with nitrogen-vacancy defects in diamond. *Reports Prog. Phys.* **77,** 056503 (2014).

31. De Sousa, R. in 183–220 (2009). doi:10.1007/978-3-540-79365-6_10

32. Jackson, J. D. *Classical electrodynamics*. (Wiley, 1999).

33. Squires, G. L. *Introduction to the Theory of Thermal Neutron Scattering*. (Cambridge University Press, 2012). doi:10.1017/CBO9781139107808

34. Kubo, R. The fluctuation-dissipation theorem. *Reports Prog. Phys.* **29,** 306 (1966).

35. Giuliani, G. *\*Quantum theory of the electron liquid.* (Cambridge University Press, 2005).

36. Van der Sar, T., Casola, F., Walsworth, R. & Yacoby, A. Nanometre-scale probing of spin waves using single-electron spins. *Nat. Commun.* **6,** 7886 (2015).

37. Agarwal, K., Schmidt, R., Halperin, B., Oganesyan, V., Zaránd, G., Lukin, M. D. & Demler, E. Magnetic noise spectroscopy as a probe of local electronic correlations in two-dimensional systems. *Phys. Rev. B* **95,** (2016).

38. Schirhagl, R., Chang, K., Loretz, M. & Degen, C. L. Nitrogen-Vacancy Centers in Diamond: Nanoscale Sensors for Physics and Biology. *Annu. Rev. Phys. Chem.* **65,** 83–105 (2014).

39. Childress, L., Walsworth, R. & Lukin, M. Atom-like crystal defects: From quantum computers to biological sensors. *Phys. Today* **67,** 38–43 (2014).

40. Jensen, K., Kehayias, P. & Budker, D. in 553–576 (Springer International Publishing, 2017). doi:10.1007/978-3-319-34070-8_18

41. Doherty, M. W., Manson, N. B., Delaney, P., Jelezko, F., Wrachtrup, J. & Hollenberg, L. C. L. The nitrogen-vacancy colour centre in diamond. *Phys. Rep.* **528,** 1–45 (2013).

42. Wrachtrup, J. & Finkler, A. Single spin magnetic resonance. *J. Magn. Reson.* **269,** 225–236 (2016).

43. Gruber, A., Dräbenstedt, A., Tietz, C., Fleury, L., Wrachtrup, J. & Borczyskowski, C. von. Scanning Confocal Optical Microscopy and Magnetic Resonance on Single Defect Centers. *Science.* **276,** 2012-2014 (1997).

44. De Lange, G., Wang, Z. H., Ristè, D., Dobrovitski, V. V & Hanson, R. Universal dynamical decoupling of a single solid-state spin from a spin bath. *Science* **330,** 60 (2010).

45. Biercuk, M. J., Uys, H., VanDevender, A. P., Shiga, N., Itano, W. M. & Bollinger, J. J. Optimized dynamical decoupling in a model quantum memory. *Nature* **458,** 996 (2009).
19

46. Bylander, J., Gustavsson, S., Yan, F., Yoshihara, F., Harrabi, K., Fitch, G., Cory, D. G., Nakamura, Y., Tsai, J.-S. & Oliver, W. D. Noise spectroscopy through dynamical decoupling with a superconducting flux qubit. *Nat. Phys.* **7,** 565–570 (2011).

47. De Lange, G., Ristè, D., Dobrovitski, V. V. & Hanson, R. Single-Spin Magnetometry with Multipulse Sensing Sequences. *Phys. Rev. Lett.* **106,** 080802 (2011).

48. Jakobi, I., Neumann, P., Wang, Y., Dasari, D. B. R., El Hallak, F., Bashir, M. A., Markham, M., Edmonds, A., Twitchen, D. & Wrachtrup, J. Measuring broadband magnetic fields on the nanoscale using a hybrid quantum register. *Nat. Nanotechnol.* (2016). doi:10.1038/nnano.2016.163

49. Stark, A., Aharon, N., Unden, T., Louzon, D., Huck, A., Retzker, A., Andersen, U. L. & Jelezko, F. Narrow-bandwidth sensing of high-frequency fields with continuous dynamical decoupling. at <https://arxiv.org/pdf/1706.04779.pdf>

50. Cai, J. M., Naydenov, B., Pfeiffer, R., McGuinness, L. P., Jahnke, K. D., Jelezko, F., Plenio, M. B. & Retzker, A. Robust dynamical decoupling with concatenated continuous driving. *New J. Phys* **14,** (2012).

51. Pham, L. M., DeVience, S. J., Casola, F., Lovchinsky, I., Sushkov, A. O., Bersin, E., Lee, J., Urbach, E., Cappellaro, P., Park, H., Yacoby, A., Lukin, M. & Walsworth, R. L. NMR technique for determining the depth of shallow nitrogen-vacancy centers in diamond. *Phys. Rev. B* **93,** 045425 (2016).

52. Rosskopf, T., Zopes, J., Boss, J. M. & Degen, C. L. A quantum spectrum analyzer enhanced by a nuclear spin memory. doi:10.1038/s41534-017-0030-6

53. Boss, J. M., Cujia, K. S., Zopes, J. & Degen, C. L. Quantum sensing with arbitrary frequency resolution. *Science* **356,** 837-840 (2017).

54. Bucher, D. B., Glenn, D. R., Lee, J., Lukin, M. D., Park, H. & Walsworth, R. L. High Resolution Magnetic Resonance Spectroscopy Using Solid-State Spins. (2017). at <http://arxiv.org/abs/1705.08887>

55. Myers, B. A., Ariyaratne, A. & Jayich, A. C. B. Double-Quantum Spin-Relaxation Limits to Coherence of Near-Surface Nitrogen-Vacancy Centers. *Phys. Rev. Lett.* **118,** 197201 (2017).

56. Momenzadeh, S. A., Stöhr, R. J., de Oliveira, F. F., Brunner, A., Denisenko, A., Yang, S., Reinhard, F. & Wrachtrup, J. Nanoengineered Diamond Waveguide as a Robust Bright Platform for Nanomagnetometry Using Shallow Nitrogen Vacancy Centers. *Nano Lett.* **15,** 165–169 (2015).

57. Siyushev, P., Kaiser, F., Jacques, V., Gerhardt, I., Bischof, S., Fedder, H., Dodson, J., Markham, M., Twitchen, D., Jelezko, F. & Wrachtrup, J. Monolithic diamond optics for single photon detection. *Appl. Phys. Lett.* **97,** 241902 (2010).

58. Riedel, D., Rohner, D., Ganzhorn, M., Kaldewey, T., Appel, P., Neu, E., Warburton, R. J. & Maletinsky, P. Low-Loss Broadband Antenna for Efficient Photon Collection from a Coherent Spin in Diamond. (2014). doi:10.1103/PhysRevApplied.2.064011

59. Tetienne, J.-P., Lombard, A., Simpson, D. A., Ritchie, C., Lu, J., Mulvaney, P. & Hollenberg, L. C. L. Scanning Nanospin Ensemble Microscope for Nanoscale Magnetic and Thermal Imaging. *Nano Lett.* **16,** 326–333 (2016).

60. Arai, K., Belthangady, C., Zhang, H., Bar-Gill, N., DeVience, S. J., Cappellaro, P., Yacoby, A. & Walsworth, R. L. Fourier magnetic imaging with nanoscale resolution and compressed sensing speed-up using electronic spins in diamond. *Nat. Nanotechnol.* **10,** 859–864 (2015).

61. Jiang, L., Hodges, J. S., Maze, J. R., Maurer, P., Taylor, J. M., Cory, D. G., Hemmer, P. R., Walsworth, R. L., Yacoby, A., Zibrov, A. S. & Lukin, M. D. Repetitive readout of a single electronic spin via quantum logic with nuclear spin ancillae. *Science* **326,** 267–72 (2009).

62. Lovchinsky, I., Sushkov, A. O., Urbach, E., de Leon, N. P., Choi, S., De Greve, K., Evans, R., Gertner, R., Bersin, E., Müller, C., McGuinness, L., Jelezko, F., Walsworth, R. L., Park, H. & Lukin, M. D. Nuclear magnetic resonance detection and spectroscopy of single proteins using quantum logic. *Science* **351,** (2016).

63. Shields, B. J., Unterreithmeier, Q. P., de Leon, N. P., Park, H. & Lukin, M. D. Efficient Readout of a Single Spin State in Diamond via Spin-to-Charge Conversion. *Phys. Rev. Lett.* **114,** 136402 (2015).

64. Jensen, K., Leefer, N., Jarmola, A., Dumeige, Y., Acosta, V. M., Kehayias, P., Patton, B. & Budker, D. Cavity-Enhanced Room-Temperature Magnetometry Using Absorption by Nitrogen-Vacancy Centers in Diamond. *Phys. Rev. Lett.* **112,** 160802 (2014).

65. Häberle, T., Oeckinghaus, T., Schmid-Lorch, D., Pfender, M., de Oliveira, F. F., Momenzadeh, S. A., Finkler, A. & Wrachtrup, J. Nuclear Quantum-Assisted Magnetometer on the Nanoscale. (2016). at <http://arxiv.org/abs/1610.03621>

66. Li, P.-B., Xiang, Z.-L., Rabl, P. & Nori, F. Hybrid quantum device with nitrogen-vacancy centers in diamond coupled to carbon nanotubes. (2016).

67. Steiner, M., Neumann, P., Beck, J., Jelezko, F. & Wrachtrup, J. Universal enhancement of the optical readout fidelity of single electron spins at nitrogen-vacancy centers in diamond. *Phys. Rev. B* **81,** 035205 (2010).

68. Wolf, T., Neumann, P., Nakamura, K., Sumiya, H., Ohshima, T., Isoya, J. & Wrachtrup, J. Subpicotesla Diamond Magnetometry. *Phys. Rev. X* **5,** 041001 (2015).

69. Barry, J. F., Turner, M. J., Schloss, J. M., Glenn, D. R., Song, Y., Lukin, M. D., Park, H. & Walsworth, R. L. Optical magnetic detection of single-neuron action potentials using quantum defects in diamond. *Proc. Natl. Acad. Sci.* **113,** 14133–14138 (2016).

70. Burek, M. J., Chu, Y., Liddy, M. S. Z., Patel, P., Rochman, J., Meesala, S., Hong, W., Quan, Q., Lukin, M. D. & Lončar, M. High quality-factor optical nanocavities in bulk single-crystal diamond. *Nat. Commun.* **5,** 5718 (2014).

71. Maletinsky, P., Hong, S., Grinolds, M. S., Hausmann, B., Lukin, M. D., Walsworth, R. L., Loncar, M. & Yacoby, A. A robust scanning diamond sensor for nanoscale imaging with single nitrogen-vacancy centres. *Nat. Nanotechnol.* **7,** 320–4 (2012).

72. Rondin, L., Tetienne, J.-P., Spinicelli, P., Dal Savio, C., Karrai, K., Dantelle, G., Thiaville, A., Rohart, S., Roch, J.-F. & Jacques, V. Nanoscale magnetic field mapping with a single spin scanning probe magnetometer. *Appl. Phys. Lett.* **100,** 153118 (2012).

73. Appel, P., Neu, E., Ganzhorn, M., Barfuss, A., Batzer, M., Gratz, M., Tschöpe, A. & Maletinsky, P. Fabrication of all diamond scanning probes for nanoscale magnetometry. *Rev. Sci. Instrum.* **87,** 063703 (2016).

74. Pelliccione, M., Myers, B. A., Pascal, L. M. A., Das, A. & Bleszynski Jayich, A. C. Two-Dimensional Nanoscale Imaging of Gadolinium Spins via Scanning Probe Relaxometry with a Single Spin in Diamond. *Phys. Rev. Appl.* **2,** 054014 (2014).

75. Häberle, T., Schmid-Lorch, D., Reinhard, F. & Wrachtrup, J. Nanoscale nuclear magnetic imaging with chemical contrast. *Nat. Nanotechnol.* **10,** 125–128 (2015).

76. Appel, P., Ganzhorn, M., Neu, E. & Maletinsky, P. Nanoscale microwave imaging with a single electron spin in diamond. *New J. Phys.* **17,** 112001 (2015).

77. Hingant, T., Tetienne, J.-P., Martínez, L. J., Garcia, K., Ravelosona, D., Roch, J.-F. & Jacques, V. Measuring the Magnetic Moment Density in Patterned Ultrathin Ferromagnets with Submicrometer Resolution. *Phys. Rev. Appl.* **4,** 014003 (2015).

78. Bonetti, S. X-ray imaging of spin currents and magnetisation dynamics at the nanoscale. *arXiv* 1611.07691 (2016). doi:10.1088/1361-648X/aa5a13

79. Tetienne, J.-P., Hingant, T., Martínez, L. J., Rohart, S., Thiaville, A., Diez, L. H., Garcia, K., Adam, J.-P., Kim, J.-V., Roch, J.-F., Miron, I. M., Gaudin, G., Vila, L., Ocker, B., Ravelosona, D. & Jacques, V. The nature of domain walls in ultrathin ferromagnets revealed by scanning nanomagnetometry. *Nat. Commun.* **6,** 6733 (2015).

80. Rondin, L., Tetienne, J.-P., Rohart, S., Thiaville, A., Hingant, T., Spinicelli, P., Roch, J.-F. & Jacques, V. Stray-field imaging of magnetic vortices with a single diamond spin. *Nat. Commun.* **4,** 2279 (2013).

81. Tetienne, J.-P., Hingant, T., Rondin, L., Rohart, S., Thiaville, A., Jué, E., Gaudin, G., Roch, J.-F. & Jacques, V. Nitrogen-vacancy-center imaging of bubble domains in a 6-Å film of cobalt with perpendicular magnetization. *J. Appl. Phys.* **115,** 17D501 (2014).

82. Tetienne, J.-P., Hingant, T., Kim, J.-V., Diez, L. H., Adam, J.-P., Garcia, K., Roch, J.-F., Rohart, S., Thiaville, A., Ravelosona, D. & Jacques, V. Nanoscale imaging and control of domain-wall hopping with a nitrogen-vacancy center microscope. *Science* **344,** 1366–9 (2014).

83. Gross, I., Martínez, L. J., Tetienne, J.-P., Hingant, T., Roch, J.-F., Garcia, K., Soucaille, R., Adam, J. P., Kim, J.-V., Rohart, S., Thiaville, A., Torrejon, J., Hayashi, M. & Jacques, V. Direct measurement of interfacial Dzyaloshinskii-Moriya interaction in X | CoFeB | MgO heterostructures with a scanning NV magnetometer ( X = Ta , TaN , and W ). *Phys. Rev. B* **94,** 064413 (2016).

84. Kosub, T., Kopte, M., Hühne, R., Appel, P., Shields, B., Maletinsky, P., Hübner, R., Liedke, M. O., Fassbender, J., Schmidt, O. G. & Makarov, D. Purely antiferromagnetic magnetoelectric random access memory. *Nat. Commun.* **8,** 13985 (2017).

85. Nagaosa, N. & Tokura, Y. Topological properties and dynamics of magnetic skyrmions. *Nat. Nanotechnol.* **8,** 899–911 (2013).

86. Shibata, K., Yu, X. Z., Hara, T., Morikawa, D., Kanazawa, N., Kimoto, K., Ishiwata, S., Matsui, Y. & Tokura, Y. Towards control of the size and helicity of skyrmions in helimagnetic alloys by spin-orbit coupling. *Nat. Nanotechnol.* **8,** 723–8 (2013).

87. Khvalkovskiy, A. V., Cros, V., Apalkov, D., Nikitin, V., Krounbi, M., Zvezdin, K. A., Anane, A., Grollier, J. & Fert, A. Matching domain-wall configuration and spin-orbit torques for efficient domain-wall motion. *Phys. Rev. B* **87,** 020402 (2013).

88. Beach, G. S. D., Tsoi, M. & Erskine, J. L. Current-induced domain wall motion. *J. Magn. Magn. Mater.* **320,** 1272–1281 (2008).

89. Parkin, S. S. P., Hayashi, M., Thomas, L., Takayanagi, H., Celestre, R. S., Church, M. M., Fakra, S., Domning, E. E., Glossinger, J. M., Kirschman, J. L., Morrison, G. Y., Plate, D. W., Smith, B. V., Warwick, T., Yashchuk, V. V., Padmore, H. A., Ustundag, E., Orenstein, J. & Ramesh, R. Magnetic domain-wall racetrack memory. *Science* **320,** 190–4 (2008).

90. Fert, A., Reyren, N. & Cros, V. Magnetic skyrmions: advances in physics and potential applications. *Nat. Rev. Mater.* **2,** 17031 (2017).

91. Boulle, O., Vogel, J., Yang, H., Pizzini, S., de Souza Chaves, D., Locatelli, A., Menteş, T. O., Sala, A., Buda-Prejbeanu, L. D., Klein, O., Belmeguenai, M., Roussigné, Y., Stashkevich, A., Chérif, S. M., Aballe, L., Foerster, M., Chshiev, M., Auffret, S., Miron, I. M. & Gaudin, G. Room-temperature chiral magnetic skyrmions in ultrathin magnetic nanostructures. *Nat. Nanotechnol.* **11,** 449–454 (2016).

92. Tetienne, J.-P., Rondin, L., Spinicelli, P., Chipaux, M., Debuisschert, T., Roch, J.-F. & Jacques, V. Magnetic-field-dependent photodynamics of single NV defects in diamond: an application to qualitative all-optical magnetic imaging. *New J. Phys.* **14,** 103033 (2012).

93. Simpson, D. A., Tetienne, J.-P., McCoey, J. M., Ganesan, K., Hall, L. T., Petrou, S., Scholten, R. E. & Hollenberg, L. C. L. Magneto-optical imaging of thin magnetic films using spins in diamond. *Sci. Rep.* **6,** 22797 (2016).

94. Hong, S., Grinolds, M. S., Pham, L. M., Le Sage, D., Luan, L., Walsworth, R. L. & Yacoby, A. Nanoscale magnetometry with NV centers in diamond. *MRS Bull.* **38,** 155–161 (2013).

95. Gould, M., Barbour, R. J., Thomas, N., Arami, H., Krishnan, K. M. & Fu, K.-M. C. Room-temperature detection of a single 19 nm super-paramagnetic nanoparticle with an imaging magnetometer. *Appl. Phys. Lett.* **105,** 072406 (2014).

96. Moriya, T. Anisotropic Superexchange Interaction and Weak Ferromagnetism. *Phys. Rev.* **120,** 91–98 (1960).

97. Gross, I., Akhtar, W., Garcia, V., Martínez, L. J., Chouaieb, S., Garcia, K., Carrétéro, C., Barthélémy, A., Appel, P., Maletinsky, P., Kim, J.-V., Chauleau, J. Y., Jaouen, N., Viret, M., Bibes, M., Fusil, S. & Jacques, V. Real-space imaging of non-collinear antiferromagnetic order with a single-spin magnetometer. *Nature* **549,** 252–256 (2017).

98. Schollwöck, U., Richter, J., Farnell, D. J. J. & Bishop, R. F. *Quantum Magnetism*. (Springer Berlin Heidelberg, 2004).

99. Furrer, A., Strässle, T. & Mesot, J. *Neutron scattering in condensed matter physics*. (World Scientific, 2009).

100. Staudacher, T., Shi, F., Pezzagna, S., Meijer, J., Du, J., Meriles, C. A., Reinhard, F. & Wrachtrup, J. Nuclear magnetic resonance spectroscopy on a (5-nanometer)3 sample volume. *Science* **339,** 561–3 (2013).

101. DeVience, S. J., Pham, L. M., Lovchinsky, I., Sushkov, A. O., Bar-Gill, N., Belthangady, C., Casola, F., Corbett, M., Zhang, H., Lukin, M., Park, H., Yacoby, A. & Walsworth, R. L. Nanoscale NMR spectroscopy and imaging of multiple nuclear species. *Nat. Nanotechnol.* **10,** 129–134 (2015).

102. Mamin, H. J., Kim, M., Sherwood, M. H., Rettner, C. T., Ohno, K., Awschalom, D. D. & Rugar, D. Nanoscale nuclear magnetic resonance with a nitrogen-vacancy spin sensor. *Science* **339,** 557–60 (2013).

103. Häberle, T., Schmid-Lorch, D., Reinhard, F. & Wrachtrup, J. Nanoscale nuclear magnetic imaging with chemical contrast. *Nat. Nanotechnol.* **10,** 125–128 (2015).

104. Müller, C., Kong, X., Cai, J.-M., Melentijević, K., Stacey, A., Markham, M., Twitchen, D., Isoya, J., Pezzagna, S., Meijer, J., Du, J. F., Plenio, M. B., Naydenov, B., McGuinness, L. P. & Jelezko, F. Nuclear magnetic resonance spectroscopy with single spin sensitivity. *Nat. Commun.* **5,** 4703 (2014).

105. Rugar, D., Mamin, H. J., Sherwood, M. H., Kim, M., Rettner, C. T., Ohno, K. & Awschalom, D. D. Proton magnetic resonance imaging using a nitrogen–vacancy spin sensor. *Nat. Nanotechnol.* **10,** 120–124 (2014).

106. Lovchinsky, I., Sanchez-Yamagishi, J. D., Urbach, E. K., Choi, S., Fang, S., Andersen, T. I., Watanabe, K., Taniguchi, T., Bylinskii, A., Kaxiras, E., Kim, P., Park, H. & Lukin, M. D. Magnetic resonance spectroscopy of an atomically thin material using a single-spin qubit. *Science* **355,** 503–507 (2017).

107. Kalinikos, B. a. & Slavin, a. N. Theory of dipole-exchange spin wave spectrum for ferromagnetic films with mixed exchange boundary conditions. *J. Phys. C Solid State Phys.* **19,** 7013–7033 (1986).

108. Farle, M. Ferromagnetic resonance of ultrathin metallic layers. *Reports Prog. Phys.* **61,** 755–826 (1998).

109. Du, C., Van der Sar, T., Zhou, T. X., Upadhyaya, P., Casola, F., Zhang, H., Ross, C. A., Walsworth, R. L., Tserkovnyak, Y. & Yacoby, A. Control and local measurement of the spin chemical potential in a magnetic insulator. *Science* **357,** 195–198 (2017).

110. Wolfe, C. S., Bhallamudi, V. P., Wang, H. L., Du, C. H., Manuilov, S., Teeling-Smith, R. M., Berger, A. J., Adur, R., Yang, F. Y. & Hammel, P. C. Off-resonant manipulation of spins in diamond via precessing magnetization of a proximal ferromagnet. *Phys. Rev. B* **89,** 180406 (2014).

111. Wolfe, C. S., Manuilov, S. A., Purser, C. M., Teeling-Smith, R., Dubs, C., Hammel, P. C. & Bhallamudi, V. P. Spatially resolved detection of complex ferromagnetic dynamics using optically detected nitrogen-vacancy spins. *Appl. Phys. Lett.* **108,** 232409 (2016).

112. Page, M. R., Guo, F., Purser, C. M., Schulze, J. G., Nakatani, T. M., Wolfe, C. S., Childress, J. R., Hammel, P. C., Fuchs, G. D. & Bhallamudi, V. P. Optically Detected Ferromagnetic Resonance in Metallic Ferromagnets via Nitrogen Vacancy Centers in Diamond. (2016). at <http://arxiv.org/abs/1607.07485>

113. Wolf, M. S., Badea, R. & Berezovsky, J. Fast nanoscale addressability of nitrogen-vacancy spins via coupling to a dynamic ferromagnetic vortex. *Nat. Commun.* **7,** 11584 (2016).

114. Andrich, P., Casas, C. F. de las, Liu, X., Bretscher, H. L., Berman, J. R., Heremans, F. J., Nealey, P. F. & Awschalom, D. D. Hybrid nanodiamond-YIG systems for efficient quantum information processing and nanoscale sensing. (2017). at <http://arxiv.org/abs/1701.07401>

115. Bauer, G. E. W., Saitoh, E. & van Wees, B. J. Spin caloritronics. *Nat. Mater.* **11,** 391–399 (2012).

116. Chen, S., Han, Z., Elahi, M. M., Habib, K. M. M., Wang, L., Wen, B., Gao, Y., Taniguchi, T., Watanabe, K., Hone, J., Ghosh, A. W. & Dean, C. R. Electron optics with p-n junctions in ballistic graphene. *Science* **353,** 1522–1525 (2016).

117. Bandurin, D. A., Torre, I., Kumar, R. K., Ben Shalom, M., Tomadin, A., Principi, A., Auton, G. H., Khestanova, E., Novoselov, K. S., Grigorieva, I. V., Ponomarenko, L. A., Geim, A. K. & Polini, M. Negative local resistance caused by viscous electron backflow in graphene. *Science* **351,** (2016).

118. Lee, M., Wallbank, J. R., Gallagher, P., Watanabe, K., Taniguchi, T., Falko, V. I. & Goldhaber-Gordon, D. Ballistic miniband conduction in a graphene superlattice. *Science* **353,** 1526–1529 (2016).

119. Young, A. F. & Kim, P. Quantum interference and Klein tunnelling in graphene heterojunctions. *Nat. Phys.* **5,** 222–226 (2009).

120. Meltzer, Alexander Y., Levin, E. & Zeldov, E. Direct Reconstruction of Two-Dimensional Currents in Thin Films from Magnetic-Field Measurements. *Phys. Rev. Appl.* **8,** 064030 (2017).

121. Tetienne, J.-P., Dontschuk, N., Broadway, D. A., Stacey, A., Simpson, D. A. & Hollenberg, L. C. L. Quantum imaging of current flow in graphene. *Sci. Adv.* **3,** (2017).

122. Waxman, A., Schlussel, Y., Groswasser, D., Acosta, V. M., Bouchard, L.-S., Budker, D. & Folman, R. Diamond magnetometry of superconducting thin films. *Phys. Rev. B* **89,** 054509 (2014).

123. Imon, S. & Ending, J. B. Local magnetic probes of superconductors. at <http://folk.uio.no/dansh/pdf/author/bending_review.pdf>

124. Clem, J. R. Theory of Flux-Flow Noise Voltage in Superconductors. *Phys. Rev. B* **1,** 2140–2155 (1970).

125. Current Distribution in Superconducting Films Carrying Quantized Fluxoids. *J. Pearl Cit. Appl. Phys. Lett* **5,** (1964).

126. Kolkowitz, S., Safira, A., High, A. A., Devlin, R. C., Choi, S., Unterreithmeier, Q. P., Patterson, D., Zibrov, A. S., Manucharyan, V. E., Park, H. & Lukin, M. D. Quantum electronics. Probing Johnson noise and ballistic transport in normal metals with a single-spin qubit. *Science* **347,** 1129–32 (2015).

127. Huang, B., Clark, G., Navarro-Moratalla, E., Klein, D. R., Cheng, R., Seyler, K. L., Zhong, D., Schmidgall, E., McGuire, M. A., Cobden, D. H., Yao, W., Xiao, D., Jarillo-Herrero, P. & Xu, X. Layer-dependent Ferromagnetism in a van der Waals Crystal down to the Monolayer Limit. (2017). at <http://arxiv.org/abs/1703.05892>

128. Gong, C., Li, L., Li, Z., Ji, H., Stern, A., Xia, Y., Cao, T., Bao, W., Wang, C., Wang, Y., Qiu, Z. Q., Cava, R. J., Louie, S. G., Xia, J. & Zhang, X. Discovery of intrinsic ferromagnetism in 2D van der Waals crystals. (2017).

129. Ma, E. Y., Cui, Y.-T., Ueda, K., Tang, S., Chen, K., Tamura, N., Wu, P. M., Fujioka, J., Tokura, Y. & Shen, Z.-X. Mobile metallic domain walls in an all-in-all-out magnetic insulator. *Science* **350,** 538–541 (2015).

130. Acosta, V. M., Bauch, E., Ledbetter, M. P., Waxman, a., Bouchard, L.-S. & Budker, D. Temperature Dependence of the Nitrogen-Vacancy Magnetic Resonance in Diamond. *Phys. Rev. Lett.* **104,** 1–4 (2010).

131. Dussaux, A., Schoenherr, P., Koumpouras, K., Chico, J., Chang, K., Lorenzelli, L., Kanazawa, N., Tokura, Y., Garst, M., Bergman, A., Degen, C. L. & Meier, D. Local dynamics of topological magnetic defects in the itinerant helimagnet FeGe. *Nat. Commun.* **7,** 12430 (2016).

132. Norman, M. R., Pines, D. & Kallin, C. The pseudogap: friend or foe of high $T_c$ ? *Adv. Phys.* **54,** 715–733 (2005).

133. Stano, P., Klinovaja, J., Yacoby, A. & Loss, D. Local spin susceptibilities of low-dimensional electron systems. *Phys. Rev. B* **88,** 045441 (2013).

134. Trifunovic, L., Pedrocchi, F. L. & Loss, D. Long-Distance Entanglement of Spin Qubits via Ferromagnet. *Phys. Rev. X* **3,** 041023 (2013).

135. Dolde, F., Doherty, M. W., Michl, J., Jakobi, I., Naydenov, B., Pezzagna, S., Meijer, J., Neumann, P., Jelezko, F., Manson, N. B. & Wrachtrup, J. Nanoscale Detection of a Single Fundamental Charge in Ambient Conditions Using the NV − Center in Diamond. *Phys. Rev. Lett.* **112,** 097603 (2014).

136. Jamonneau, P., Lesik, M., Tetienne, J. P., Alvizu, I., Mayer, L., Dréau, A., Kosen, S., Roch, J.-F., Pezzagna, S., Meijer, J., Teraji, T., Kubo, Y., Bertet, P., Maze, J. R. & Jacques, V. Competition between electric field and magnetic field noise in the decoherence of a single spin in diamond. *Phys. Rev. B* **93,** 024305 (2016).

137. Tisler, J., Oeckinghaus, T., Stöhr, R. J., Kolesov, R., Reuter, R., Reinhard, F. & Wrachtrup, J. Single Defect Center Scanning Near-Field Optical Microscopy on Graphene. *Nano Lett.* **13,** 3152–3156 (2013).

138. Brenneis, A., Gaudreau, L., Seifert, M., Karl, H., Brandt, M. S., Huebl, H., Garrido, J. A., Koppens, F. H. L. & Holleitner, A. W. Ultrafast electronic readout of diamond nitrogen–vacancy centres coupled to graphene. *Nat. Nanotechnol.* **10,** 135–139 (2014).

139. Kézsmárki, I., Bordács, S., Milde, P., Neuber, E., Eng, L. M., White, J. S., Rønnow, H. M., Dewhurst, C. D., Mochizuki, M., Yanai, K., Nakamura, H., Ehlers, D., Tsurkan, V. & Loidl, a. Néel-type skyrmion lattice with confined orientation in the polar magnetic semiconductor GaV4S8. *Nat. Mater.* **14,** 1116–1122 (2015).

140. West, A. D., Hayward, T. J., Weatherill, K. J., Schrefl, T., Allwood, D. a & Hughes, I. G. A simple model for calculating magnetic nanowire domain wall fringing fields. *J. Phys. D Appl. Phys.* **45,** 095002 (2012).